\documentclass[pdflatex,sn-nature]{sn-jnl}

\usepackage{graphicx}%
\usepackage{multirow}%
\usepackage{amsmath,amssymb,amsfonts}%
\usepackage{amsthm}%
\usepackage{mathrsfs}%
\usepackage[title]{appendix}%
\usepackage{xcolor}%
\usepackage{color,soul}
\usepackage{textcomp}%
\usepackage{manyfoot}%
\usepackage{booktabs}%
\usepackage{algorithm}%
\usepackage{algorithmicx}%
\usepackage{algpseudocode}%
\usepackage{listings}%
\usepackage[english]{babel}
\usepackage{bm}
\usepackage{indentfirst}
\usepackage{hyperref}

\begin{document}

\title[Short spin wave excitation in nanophotonic structures]{Magnetophotonic waveguide nanostructure for selective ultrafast optical excitation of high-K spin dynamics}


\author*[1,2]{\fnm{Lutsenko} \sur{Savelii V.}}\email{savlucenko@yandex.ru}

\author[1]{\fnm{Khramova} \sur{Anastasia E.}}\email{ae.khramova@gmail.com}

\author[1,2]{\fnm{Ignatyeva} \sur{Daria O.}}\email{daria.ignatyeva@gmail.com}

\author[2]{\fnm{Konkov} \sur{Daniil V.}}\email{konkov.dv19@physics.msu.ru}

\author[3]{\fnm{Kaurova} \sur{Natalia S.}}\email{nkaurova@yandex.ru}

\author[4]{\fnm{Syrov} \sur{Anatoly A.}}\email{anatoly199824@rambler.ru}

\author[4]{\fnm{Kudryashov} \sur{Alexander L.}}\email{alex.kudryashov.leo@mail.ru}

\author[3]{\fnm{Goltsman} \sur{Grigory N.}}\email{goltsman10@mail.ru}

\author[4]{\fnm{Berzhansky} \sur{Vladimir N.}}\email{v.n.berzhansky@gmail.com}

\author[1,2]{\fnm{Belotelov} \sur{Vladimir I.}}\email{v.belotelov@rqc.ru}

\equalcont{These authors contributed equally to this work.}

\affil*[1]{\orgname{Russian Quantum Center}, \orgaddress{\city{Moscow}, \postcode{121205}, \country{Russia}}}

\affil[2]{\orgdiv{Faculty of Physics}, \orgname{Lomonosov Moscow State University}, \orgaddress{\street{Leninskie Gory}, \city{Moscow}, \postcode{119991}, \country{Russia}}}

\affil[3]{\orgname{Moscow State Pedagogical University}, \orgaddress{\street{Malaya Pirogovskaya}, \city{Moscow}, \postcode{119435}, \country{Russia}}}

\affil[4]{\orgname{V.I. Vernadsky Crimean Federal University}, \orgaddress{\street{Vernadsky Ave. 4}, \city{Simferopol}, \postcode{295007}, \country{Russia}}}


\abstract{Optomagnonics provides a promising method of a Joule-loss-free spin control that can be performed at ultrafast timescale. However, the cornerstone of optomagnonics is impossibility to focus the light tighter than a diffraction limit. This constrains minimal size of the optically switched magnetic bits and minimal wavelengths of the optically excited spin waves in optomagnonic devices thus preventing its further progress. Here we propose and experimentally demonstrate a novel method of the selective optical excitation of the short exchange spin waves by using a specially designed magnetophotonic grating. This method is based on the creation of the sign-changing profile of the inverse Faraday effect (IFE) induced in a magnetic film due to excitation of the optical guided TE-mode by a femtosecond laser pulse. The spatial period of the IFE profile is subdiffractive which allows to launch narrow band short spin waves whose wavelengths are around 300~nm as was experimentally demonstrated and potentially can be made down to $\sim 100$~nm. This opens new horizons for optomagnetic applications that are wide ranging, from logical elements to data processing devices.}


\maketitle

\section*{Introduction}\label{sec1}

In the last decade optomagnonics dealing with optical switching and excitation of spin waves in magnets has been taken much research interest. It is mainly due to its potential applications for low dissipative ultrafast data storage and data processing devices which are advantageous with respect to contemporary semiconductor electronics and conventional data storage by electric pulse generated bias magnetic fields ~\cite{kimel20222022,kalashnikova2015ultrafast, kimel2007femtosecond}. Optomagnonic information processing devices~\cite{kajiwara2010transmission,chen2021reconfigurable}, logical elements~\cite{kolosvetov2022concept,jamali2013spin}, memory~\cite{kimel2019writing,stanciu2007all,gundougan2015solid,stupakiewicz2019selection,stupakiewicz2017ultrafast}, and other spin-based devices~\cite{nikitov2015magnonics} were recently demonstrated. Information in optomagnonic devices is stored as the spin states, and is written~\cite{stupakiewicz2017ultrafast,le2015nanoscale,stupakiewicz2019selection,cheng2020all,im2019all} and processed by optical pulses via various optomagnetic effects~\cite{kudlis2021all,kirilyuk2010ultrafast, khokhlov2019optical, frej2023laser, zhu2022inverse,shen2018dominant}. Spins as the information carriers make the optomagnonic devices free of Joule losses. On the other hand, optical spin control via the femtosecond laser pulses can be performed on the picosecond time scales~\cite{mashkovich2021terahertz} that allows to increase the information processing speeds up to THz. Optomagnonics provides wide possibilities to tune the parameters of the optical impact on the system by choosing the parameters of the femtosecond pulse and its position~\cite{khramova2023tuning,yoshimine2017unidirectional,filatov2024tunable}. This is a key feature for the local spin manipulation~\cite{ignatyeva2019plasmonic, ignatyeva2024optical,dutta2017surface} and selectivity~\cite{chernov2020all,krichevsky2021selective, krichevsky2024spatially}. However, the key problem of the optomagnonics is a quite large area addressed by the femtosecond pulse.

Nonthermal optomagnetic effects launch spin dynamics in a part of a sample which is illuminated by the pump laser pulse acting on spins during a few hundreds of femtoseconds by a torque whose magnitude is proportional to the light intensity and the phase homogeneous across the beam area. As the optical pulse duration is much shorter than a period of spin dynamics the optically induced torque acts on spins as a kind of photonic kick. Spatial distribution of such photonic kick has Gaussian character and its minimal area is constrained in the lateral direction by the Rayleigh diffraction limit. As a result, in lateral direction a wide spectrum of the spin waves with the wavelengths ranging from infinity down to about doubled diameter~\cite{savochkin2017generation} is excited. It leads to rather fast dephasing of different spin harmonics at a time scale of a few nanoseconds even in low dissipative dielectric magnets and, on the other hand, prohibits excitation of short spin waves whose wavelengths are at the submicron scales. Tight focusing of the femtosecond pulse allows for generation of the shorter spin waves. However, an ability to focus the light is theoretically limited by a diffraction limit and practically limited by the experimental construction of the pump-probe setups. The shortest spin waves excited by the laser pulses in the magnetic films have still been of several micrometer values~\cite{khramova2023tuning, hashimoto2017all}. It significantly limits capabilities of the ultrafast optical means for spin waves generation.

As for the direction longitudinal to the laser pulse propagation, in the transparency spectral region of a magnet the pulse penetrates through the entire sample providing almost uniform photonic kick in-depth. If the pump is strongly absorbed in the magnet then it excites spins only in the near surface parts of the sample which allows launching spin waves propagating in-depth of the magnetic crystal ~\cite{afanasiev2021ultrafast}. However, it should be noted that such mechanism is intrinsically thermal as it is based on the light absorption which is not favorable for applications. Additionally, the spectrum of the excited spin waves though contains short spin waves still remains rather wide.

Coming back to the nonthermal optomagnonic effects necessary light localization in depth can be achieved even in the transparency spectral region using various magnetophotonic nanostructures~\cite{qin2022nanophotonic}, such as plasmonic~\cite{yang2022observation, floess2018nonreciprocal,novikov2020ultrafast,armelles2013magnetoplasmonics, maccaferri2015resonant, garoli2019plasmonic} or all-dielectric~\cite{bsawmaii2023large,ignatyeva2022all, almpanis2018dielectric, smyrnakis2015optical, pantazopoulos2017photomagnonic} that help localizing light in a magnetic material and overcome these limits. Thus, it was experimentally shown that guided mode gratings~\cite{chernov2020all}, slab waveguides~\cite{krichevsky2021selective}, magnetophotonic crystals~\cite{krichevsky2024spatially} produce inhomogeneous electromagnetic field across the film thickness and launch standing spin waves with the wavelengths down to $\sim 50$~nm. 

Nevertheless, the problem of nonthermal optical excitation of propagating short spin waves still remains unsolved. 

At the same time, short spin waves are of great practical interest~\cite{gross2020building}. The group velocity of the spin waves increases with a decrease of its wavelength due to the exchange contribution~\cite{kalinikos1990dipole,shiota2020observation}. Large group velocities of the short spin waves make the signal transmission faster. Moreover, the size of the optomagnonic device is limited by the wavelength of the spin wave used in it, so that excitation of the short spin waves enables miniaturization of the devices. 

Such short-wavelength spin waves were recently excited by microwave fields~\cite{demidov2011excitation,wang2023deeply,yu2016approaching,liu2018long}. As the wavelength of the spin wave is limited by the microwave antenna size~\cite{ciubotaru2016all}, special methods are developed to overcome this limitation. Wang et al.~\cite{wang2023deeply} proposed a nonlinear mechanism of the conversion of the ferromagnetic resonance mode to the propagating spin wave with a wavelength up to 200~nm. Yu et al.~\cite{yu2016approaching} demonstrated excitation of spin waves with wavelengths down to 88~nm using the microwave-to-magnon transducers in the form of the magnetic nanodisks placed on the top of the iron-garnet film. Spin waves with the wavelengths of 50~nm were excited in a hybrid system of ferrimagnetic nanowires on the top of the iron-garnet film in the work by Liu et al.~\cite{liu2018long}.

However, in turn, microwave magnon excitation suffer from various practical disadvantages. The speed of such devices is limited by the frequency of the microwave field. Actually, it coincides with the limits of the electronic devices producing the electric currents in the antennas while optomagnonics allows for the ultrafast up to THz operating rates. Also, it is important that optomagnetism allows for on-the-fly tunability, as it was mentioned above. Additionally, optical means benefit from other kind of tunability including control of spectrum of the spin excitations by a simple readjustments of the pump wavelength, angle of incidence or polarization.

Here we propose and experimentally demonstrate a novel nonthermal method of the selective optical generation and observation of the short spin waves in magnetophotonic waveguide gratings. This method is based on the creation of the sign-changing pattern of the inverse Faraday effect (IFE) field using the excitation of the optical resonances in the magnetophotonic structure. The spin waves launched in such a device have wavelengths corresponding to the the IFE distribution periods. Short spin waves with the wavelengths as small as 300~nm were experimentally launched. This method allows for the selective excitation of spin waves with wavelengths down to $\sim100$nm. Moreover, the spin-wave wavelength can be tuned by the waveguide grating parameters or by the wavelength and angle of incidence of the optical pulse. The exchange contribution becomes important for such short spin waves, making the group velocities high.

\section*{Peculiar spin dynamics in a magnetophotonic waveguide grating}\label{subsec3}

Optomagnetic laser-induced inverse Faraday effect is described in terms of the effective magnetic field:
\begin{equation}
\mathbf{H^\mathrm{IFE}} = -\frac{g \epsilon_0}{4M_s} \mathrm{Im} [\mathbf{E \times E^*}],
\label{Eq: IFE}
\end{equation}
where $g$ is the gyration coefficient and $M_s$ is the saturation magnetization of the magnetic film. It produces an ultrashort torque $\textbf{T} = [\mathbf{M \times\mathbf{H^\mathrm{IFE}}}]$ acting on magnetic moments and launching spin dynamics in the area where the pump laser pulse is focused. The magnitude of this torque is proportional to the light intensity, and the phase is homogeneous across the whole beam area. Since laser pulse duration is much shorter than the period of magnetization precession the torque can be considered instantaneous and it determines initial conditions for the excited spin dynamics. On the other hand, spatial distribution of $\mathbf{H^\mathrm{IFE}}(\textbf{r})$ and $\textbf{T}(\textbf{r})$ determines spatial and therefore frequency spectrum of the excited spin waves. If in the equilibrium state the magnetization is in-plane and along y-axis, then the IFE field of the normally incident pump provides torque acting on magnetization in x-axis direction. 
For the pump beam of Gaussian shape the Fourier transform of $T_x(\textbf{r})$  in a smooth magnetic film is given by $\hat T_x(\mathbf K)=T_0r_0 \sqrt \pi \exp(-K^2 r_0^2/4)$, where $T_0$ is maximum value of the torque and $r_0$ is radius of the beam . It allows to estimate wavenumber $K$ bandwidth of the excited spin waves from $K=0$ to $K_max=2/r_0$~\cite{savochkin2017generation}. Tight focusing of the pump beam to the spots with a smaller $r_0$ allows for generation of the shorter spin waves. However, an ability to focus the light is constrained by a diffraction limit and practically limited by the numerical aperture of the focusing objective. So far the narrowest laser spot exciting the spin dynamics in the magnetic films was still several micrometers wide and therefore the spin-wave wavenumbers were up to $K\approx 1\cdot 10^4~\mathrm{cm}^{-1}$.

\begin{figure}[htbp]
    \centering
    \textbf{a}\\
    \includegraphics[width=0.5\linewidth]{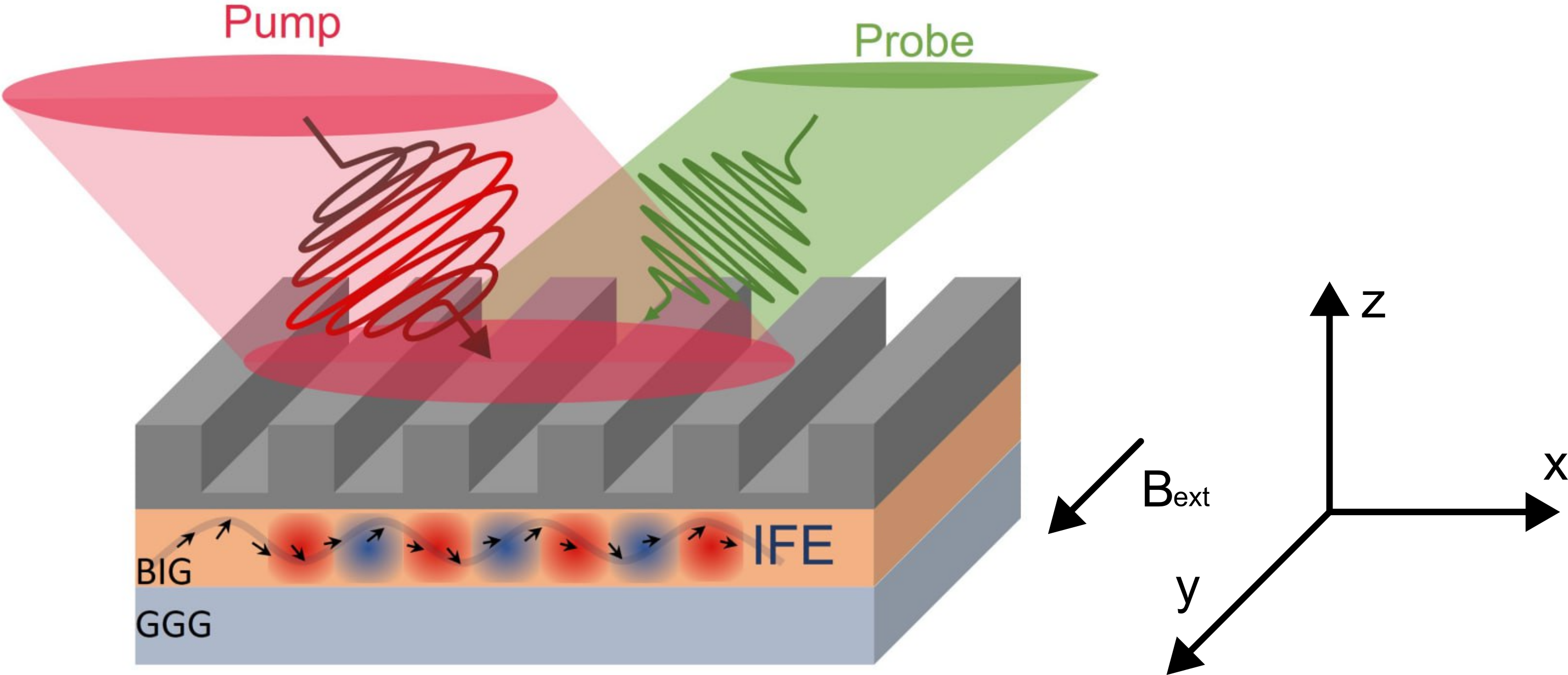}\\
    \textbf{b}~~~~~~~~~~~~~~~~~~~~~~~~~~~~~~~~~~~~~~~~~~~~~~~~~~~~~~~\textbf{c}\\
    \includegraphics[width=0.48\linewidth]{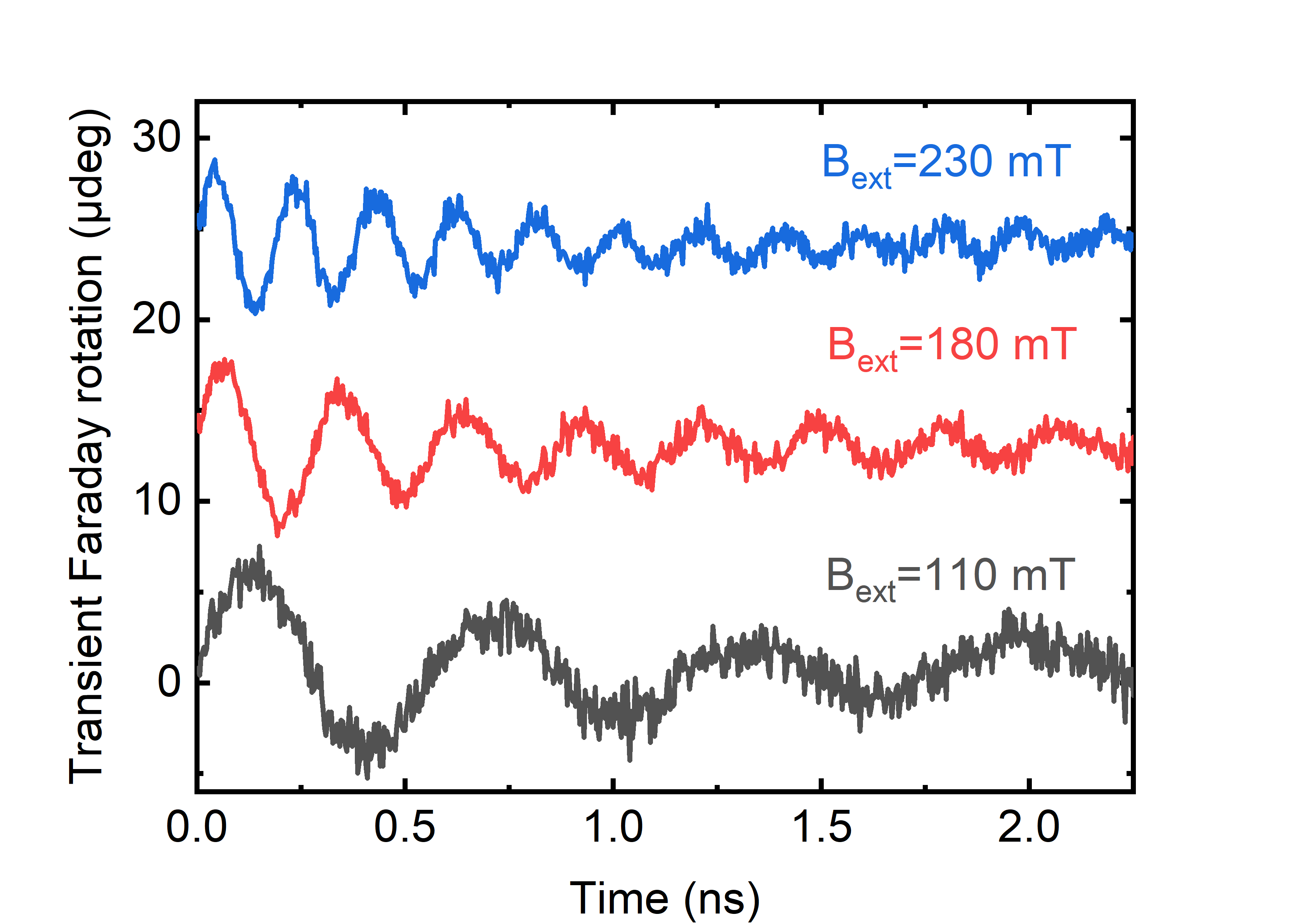}
    \includegraphics[width=0.48\linewidth]{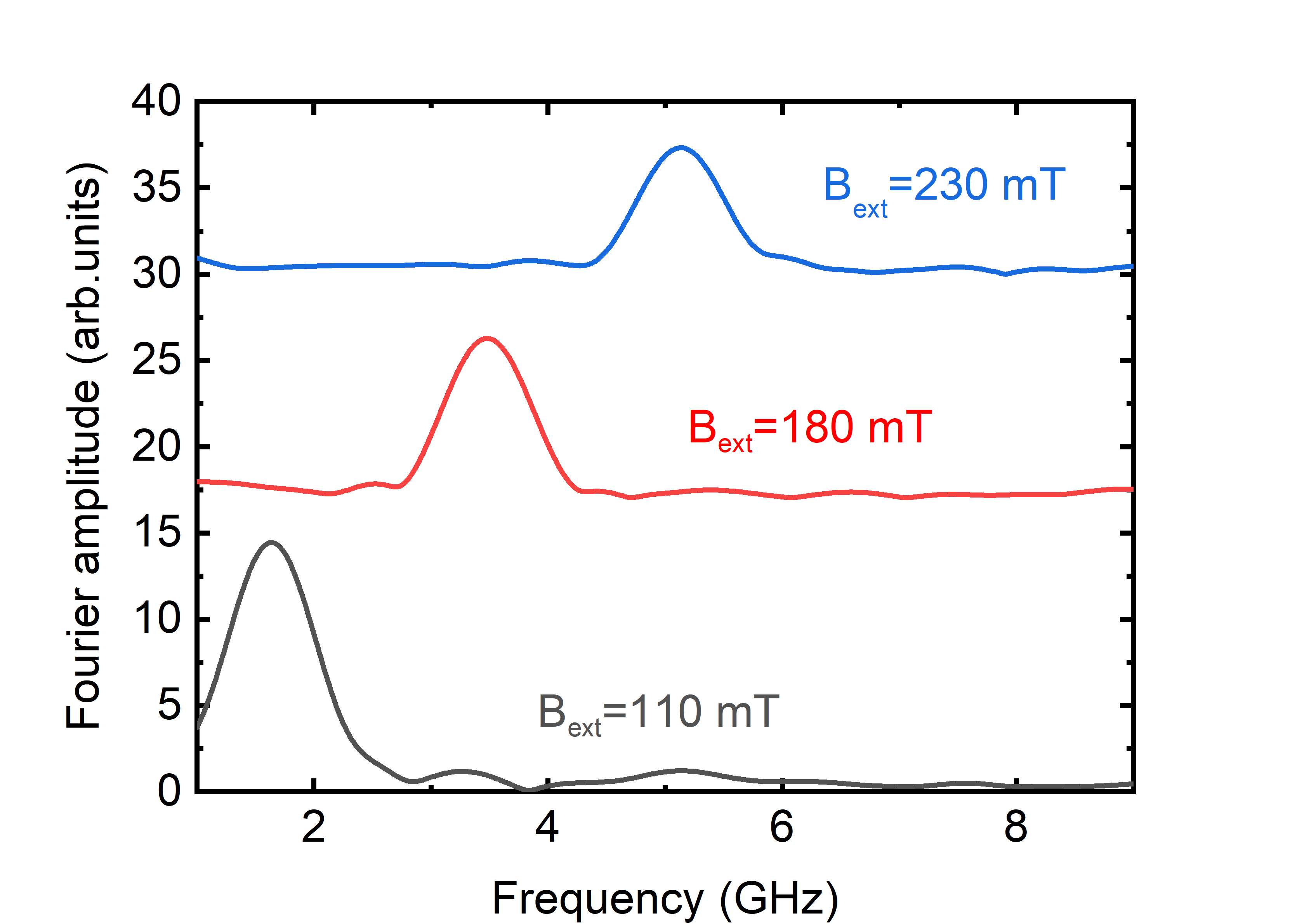}\\    \textbf{d}~~~~~~~~~~~~~~~~~~~~~~~~~~~~~~~~~~~~~~~~~~~~~~~~~~~~~~~\textbf{e}\\
    \includegraphics[width=0.48\linewidth]{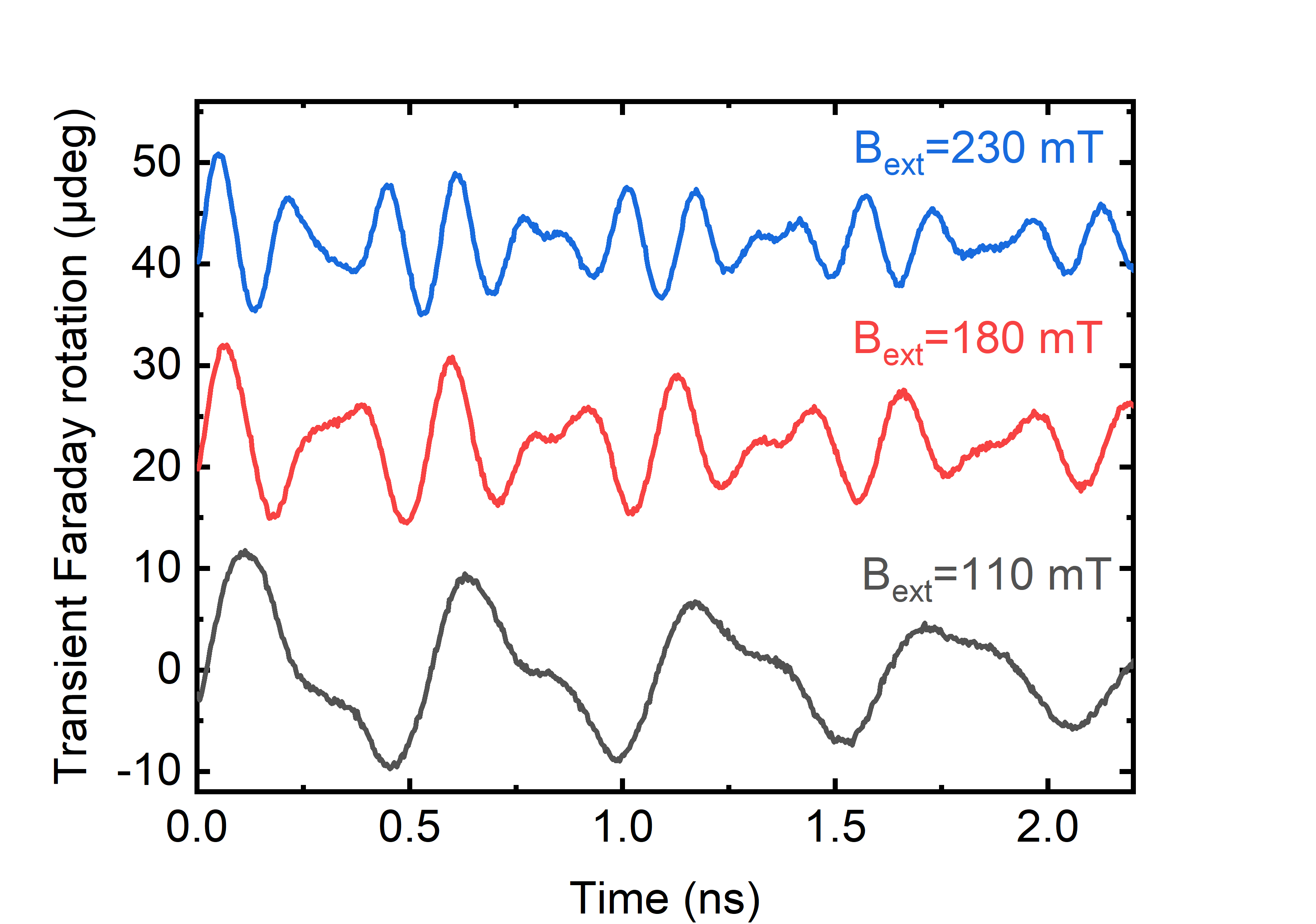}
    \includegraphics[width=0.48\linewidth]{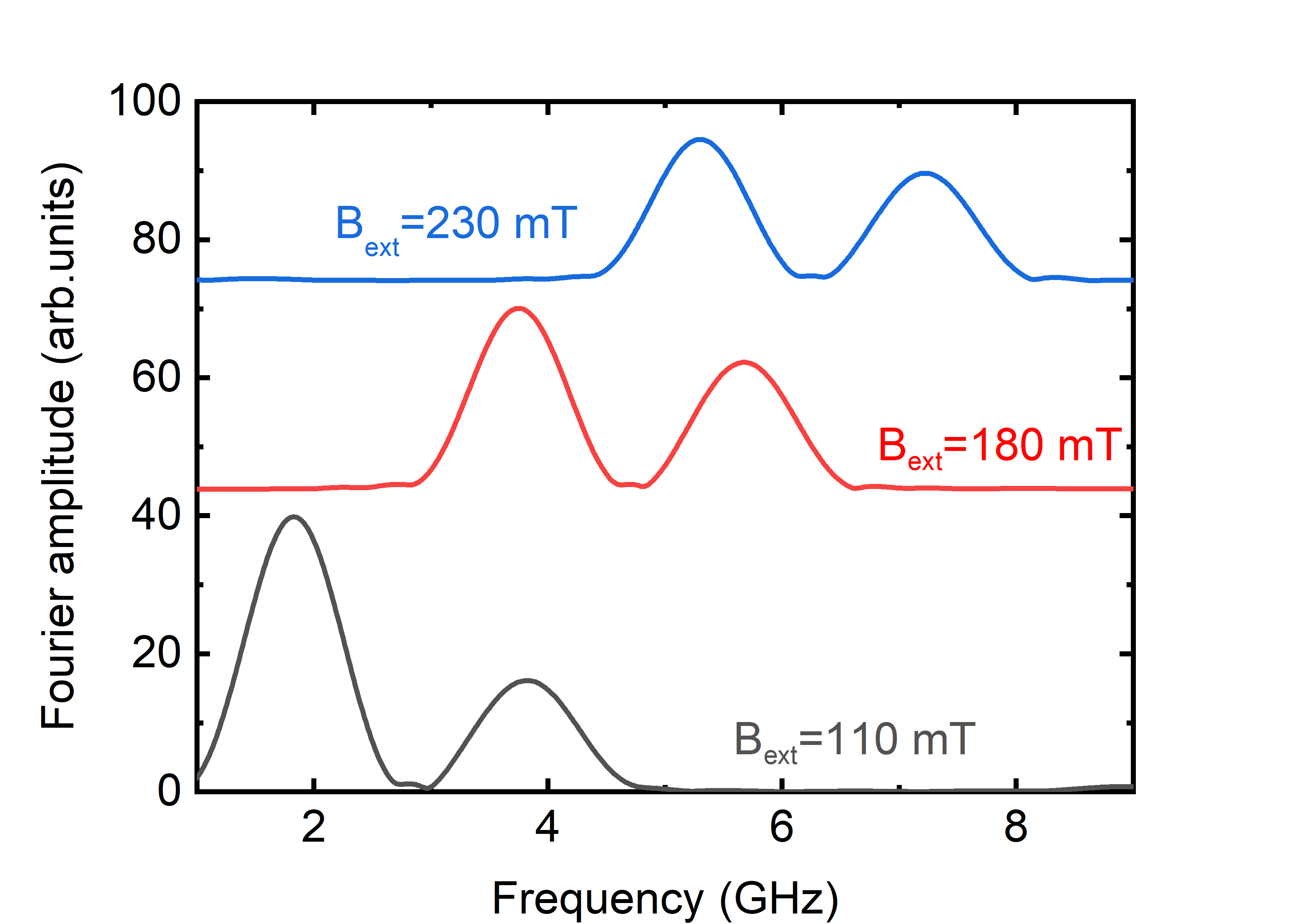}
    \caption{\textbf{a} Scheme of the short spin wave excitation via a nanophotonic waveguide grating. \textbf{b,d} Spin precession and \textbf{c,e} its Fourier spectra excited in a uncovered BIG film (b,c) and in the waveguide grating fabricated on the basis of the same film (d,e). The measurements were performed with a 685-nm femtosecond laser pump pulse of circular polarization for several values of the external in-plane magnetic field from 110 mT to 230 mT.}
    \label{Fig: Scheme and precession}
\end{figure}

The diffraction obstacle might be overcome if $\textbf{T}(\textbf{r})$ distribution of Gaussian shape is additionally modulated at much shorter scales. This problem can be solved if some kind of optical mode is excited inside the magnetic film. It can be implemented, for example, by a guided mode. For that reason here we consider a magnetic film covered by a nonmagnetic grating. In particular, the grating was made by 1D nanopatterning of $d_\mathrm{TiO_2}=130$~nm thick $\mathrm{TiO_2}$ layer on top of the $d=95$~nm thick BIG film. Since refractive index of the BIG film is larger than that of the substrate the entire structure represents a kind of magnetophotonic waveguide grating (MPWG) (see Fig.~\ref{Fig: Scheme and precession}a).

For the experimental studies we use a circularly polarized femtosecond laser pulse focused in a beam of radius $r_0=3.5~\mu\mathrm{m}$ (see Methods for details) to excite spin dynamics in a 95-nm thick bismuth-substituted iron-garnet (BIG) {$\mathrm{(Bi,Y,Lu)_3(Fe,Ga)_5O_{12}}$} ferrimagnetic film. The excited spin precession is measured by a transient Faraday rotation of the probe pulse coming at some variable delay with respect to the pump pulse. Expectedly, the probe signal has a decaying harmonics shape (Fig.~\ref{Fig: Scheme and precession}b). Its Fourier spectrum demonstrates one pronounced peak at the corresponding frequency (Fig.~\ref{Fig: Scheme and precession}c) that shifts towards higher frequencies as the external magnetic field increases. While the detailed analysis of this signal will be provided further, let us underline that it is a typical picture for the optical excitation of spin waves with long wavelengths and $K\approx 0$.

The situation drastically changes if the magnetic film is covered by a nonmagnetic grating. In particular, the grating was made by 1D nanopatterning of 130-nm-thick $\mathrm{TiO_2}$ layer on top of the BIG film. Since refractive index of the BIG film is larger than that of the substrate the entire structure represents a kind of magnetophotonic waveguide grating (MPWG) (see Fig.~\ref{Fig: Scheme and precession}a). Spin dynamics in this grating is launched via the same circularly polarized pump femtosecond pulse of $r_0=3.5~\mu\mathrm{m}$ radius and 685-nm wavelength which is resonant for the magnetophotonic structure if the beam is incident at 17 deg. 

Figure~\ref{Fig: Scheme and precession} shows a prominent difference of the spin precession observed in a smooth BIG film (b,c) and in the magnetophotonic grating fabricated based on the same magnetic film (d,e). In the latter case the observed spin dynamics exists for much longer times and has a quite unusual periodic beating (Fig.~\ref{Fig: Scheme and precession}d). Its Fourier transform has two maxima both of which exhibit similar frequency growth with an increase of the external magnetic field (Fig.~\ref{Fig: Scheme and precession}e). The peak with the lower frequency coincides with the one observed in the bare film and corresponds to the excitation of the long-wavelength spin waves with $K\approx 0$. On the other hand, the higher frequency peak hints to the excitation of spin waves with a much shorter wavelength. Let us analyze its origin and character. 

\section*{Distribution of the optically induced magnetic torque}\label{subsec4}

\begin{figure}[ht]
    \centering
    \textbf{a}~~~~~~~~~~~~~~~~~~~~~~~~~~~~~~~~~~~~~~~\textbf{b}\\
    \includegraphics[width=0.35\linewidth]{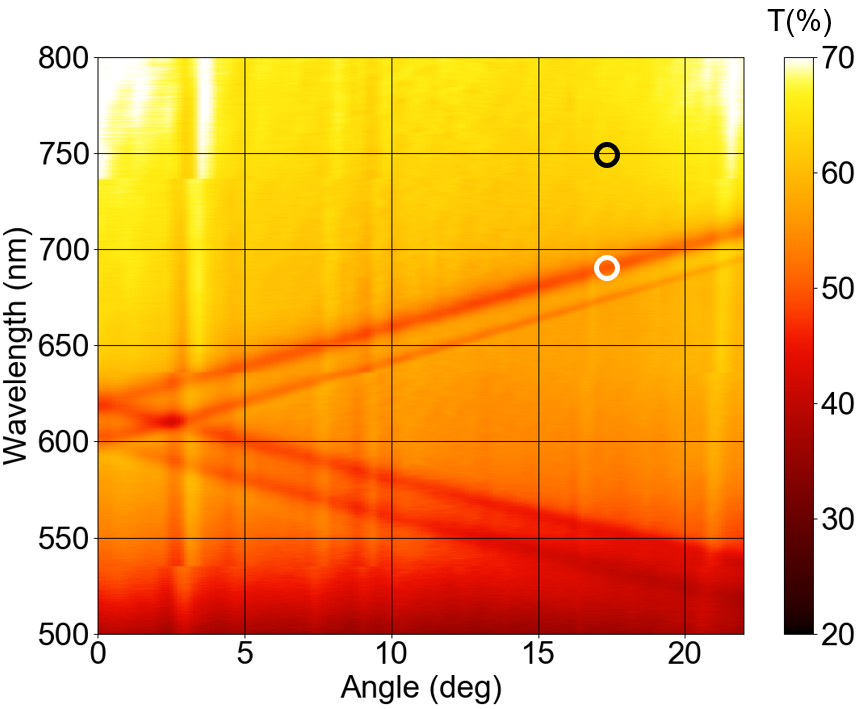}
    \includegraphics[width=0.35\linewidth]{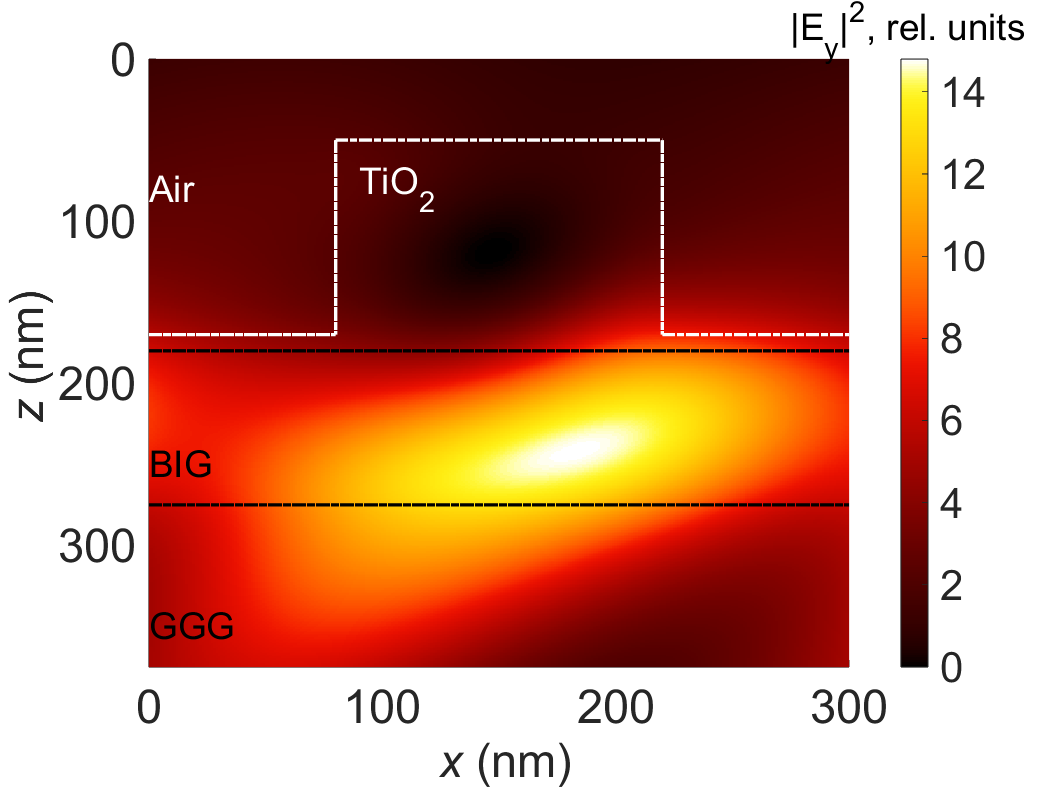}\\
 \textbf{c}~~~~~~~~~~~~~~~~~~~~~~~~~~~~~~~~~~~~~~~\textbf{d}\\
 \includegraphics[width=0.35\linewidth]{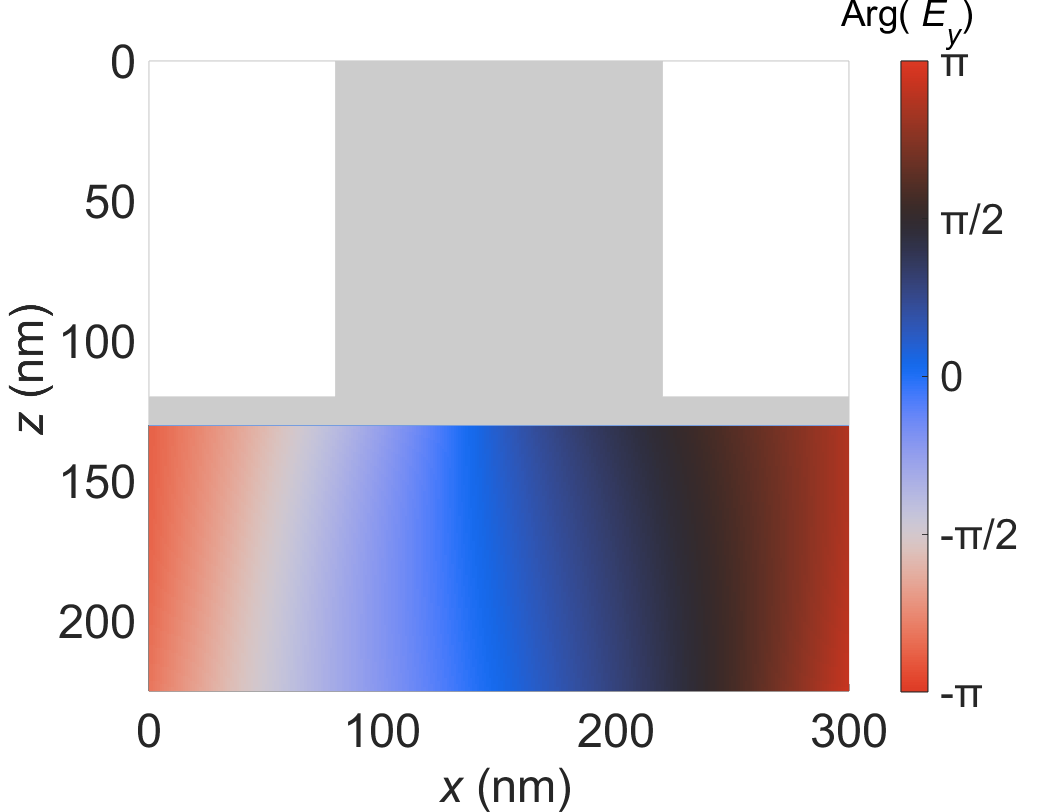}
    \includegraphics[width=0.35\linewidth]{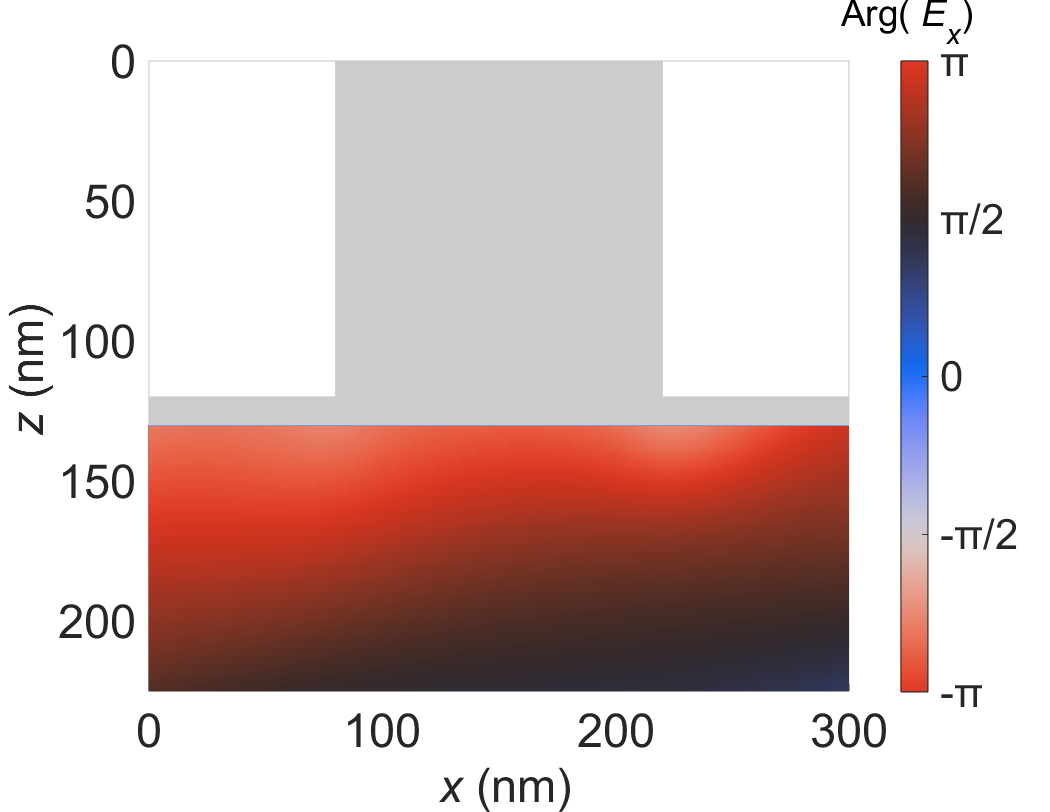}
    \\
\textbf{e}~~~~~~~~~~~~~~~~~~~~~~~~~~~~~~~~~~~~~~~\textbf{f}\\
    \includegraphics[width=0.33\linewidth]{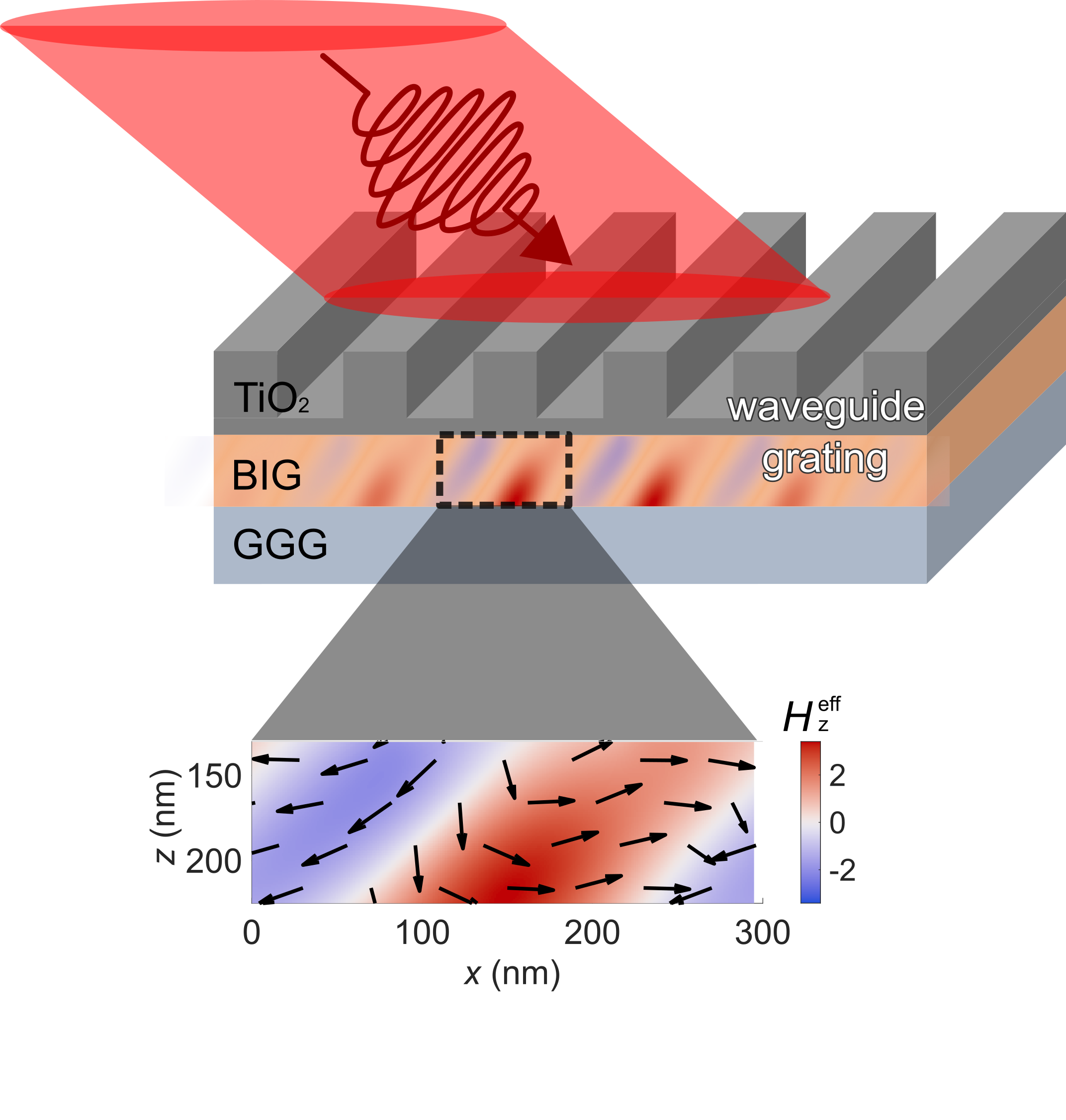}
    \includegraphics[width=0.33\linewidth]{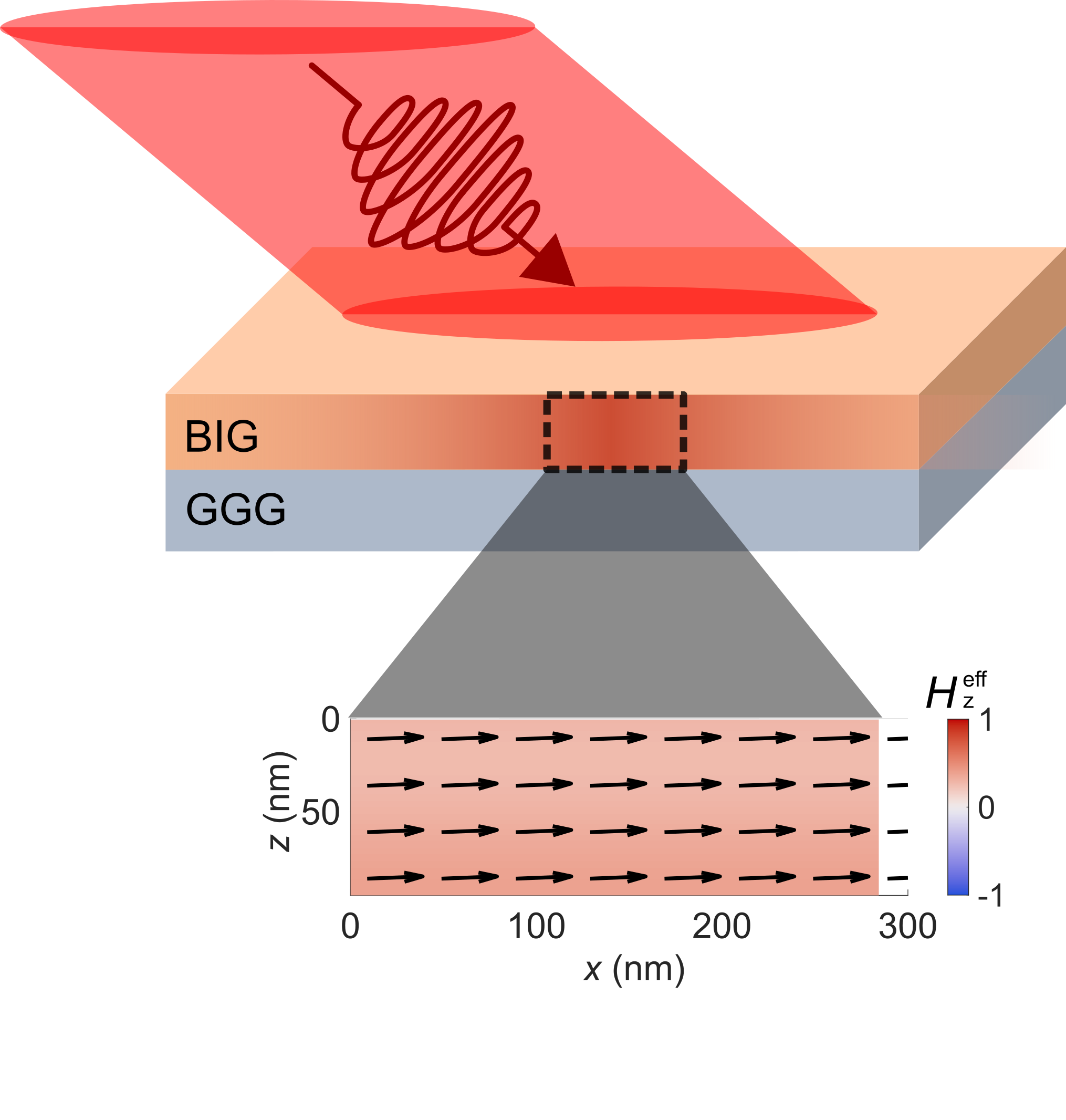}
    
    \caption{\textbf{a} The transmittance spectra of the magnetophotonic waveguide grating measured for the circularly polarized incident light. White and black circles show the wavelengths used in experiment as in- and out-of-resonance pump, respectively. \textbf{b-f} Calculated distribution of the TE-mode intensity represented by $|E_y|^2$ (\textbf{b}), of phases of the optical field components $E_y$ (\textbf{c}) and $E_x$ (\textbf{d}) and out-of-plane component of the IFE effect effective magnetic field $H^\mathrm{IFE}_z$ in the MPWG (\textbf{e}) and in the same bare film (\textbf{f}). Arrows in (\textbf{e,f}) show distribution of the IFE induced torque on magnetization (\textbf{T}). The calculations are performed for the oblique incidence at 17 deg. of the circularly polarized light at $\lambda=685$~nm which corresponds to the TE-mode resonance. } 
    \label{Fig: Optical properties}
\end{figure}

The magnetophotonic structure was designed to support excitation of the guided modes. The transmission spectrum of MPWG with $P=300$~nm and 140-nm wide stripes was measured for the circularly polarized incident light in a wide angular and wavelength range (Fig.~\ref{Fig: Optical properties}a). The spectrum demonstrates excitation of the TE and TM guided modes by $m=\pm1$ grating diffraction order. They are clearly seen as the dips in the transmittance whose spectral position is described by a dispersion~\cite{voronov2020magneto}:
\begin{equation}
k_0 \sin{\theta}+2\pi\frac{m}{P}  = \pm n_\mathrm{wg} k_0,
\label{Eq: WG disp}
\end{equation}
where $k_0 =2\pi/\lambda$ is a light wavevector, $\theta$ is the angle of incidence, $P$ is a grating period, $n_\mathrm{wg}$ is a refractive index of the guided mode, and $m \in \mathbb{Z}$ is an integer denoting the number of the grating diffraction order that excites the mode. The type of mode, TE or TM, is distinguished by the measurements of the similar spectra with a linearly polarized incident light (see Methods). At the normal incidence $\theta=0^\circ$, modes propagating in $+x$ and $-x$ directions are excited simultaneously, forming a standing wave inside MPWG. We use an oblique incidence of the pump pulse at an angle of $\theta=17^\circ$. It allows to excite a TE-mode propagating only in $-x$ direction (Fig.~\ref{Fig: Optical properties}b) at a wavelength of $\lambda=685~nm$ by $m=1$ (Fig.~\ref{Fig: Optical properties}a, white circle). The choice of a specific 17-degree light incidence angle is caused by the pump-probe setup (see Methods). Though the TE-mode is mainly localized in the magnetic film (Fig.~\ref{Fig: Optical properties}b), the upper $\mathrm{TiO_2}$ grating influences its properties.    
As the inverse Faraday effect is very sensitive to the polarization of light inside a magnetic medium (see Eq.~\ref{Eq: IFE}) and the magnetophotonic grating significantly changes the light polarization inside the magnetic film due to the excitation of the optical modes, it is important to analyze both the amplitude and the phase distributions of the electromagnetic field components inside the magnetic material. The calculations were performed using the rigorous coupled wave analysis method (RCWA, see Methods for details).  

Illumination of the sample with circularly polarized light of wavelength $\lambda=685~nm$ at slightly oblique incidence provides two main components of the optical field inside the MPWG: $E_x$ and $E_y$ which have quite different character. Indeed, at $\lambda=685~nm$ and 17 deg. angle of incidence the TE-mode is excited and its excitation is accompanied by the $E_y$ phase profile typical for the propagating TE-mode(Fig.~\ref{Fig: Optical properties}c). The phase of $E_y$ is almost constant along $z$-axis and gradually decreases along $x$-axis by $2\pi$ on the scale of one grating period which confirms the TE mode propagation in $-x$ direction. On the contrary, as TM-mode is not excited at $\lambda=685$~nm (its resonance is at $\lambda=670$~nm), the phase of the $E_x$ component of light has a non-resonant character and is nearly constant along $x$-axis except for a small variation caused by the oblique incidence and the presence of the small tangential wavevector component (Fig.~\ref{Fig: Optical properties}d). The resulting helicity of light in the magnetic film is determined by the relative phases between the $E_x$ and $E_y$ components. Thus, it changes from right to left circular polarization on the scale of one period of a waveguide grating. As a result, an out-of-plane component of the effective magnetic field induced via the inverse Faraday effect $H^\mathrm{IFE}_z = -\frac{ g \epsilon_0}{4M_s} \mathrm{Im}(E_xE_y^*-E_yE_x^*)$ becomes sign-changing with a relatively small period equal to the grating period 300~nm (see red-blue color scheme in Fig.~\ref{Fig: Optical properties}e). Such distribution of $H^\mathrm{IFE}_z$ provides periodic in space ultrashort (250 fs in duration) torque $\textbf{T} = [\mathbf{M \times\mathbf{H^\mathrm{IFE}}}]$ acting on spins mostly along x-axis one half of the spatial period (150 nm), and opposite to x-axis the second half of the spatial period (150 nm) (see black arrows in Fig.~\ref{Fig: Optical properties}e). It represents quite a difference with conventional situation when circularly polarized light propagates through the magnetic film and induces in it a uniformly distributed $\mathbf{H^\mathrm{IFE}}$ acting on spins instantaneously with a torque $\textbf{T}$ directed along x-axis all over the pump beam (see black arrows in Fig.~\ref{Fig: Optical properties}f). One should also notice one order of magnitude enhancement of the IFE on the guided mode with respect to the uniform excitation (compare Fig.~\ref{Fig: Optical properties}e and f).    

Therefore, the physical origin of such a sign-changing IFE behavior in the MPWG is excitation of the TE-guided mode. Notice that the opposite situation, excitation of the TM-guided mode is not similar to the discussed one. As $E_x$ component of the TM-mode is sign-changing in depth even for the $0^\mathrm{th}$-order in depth mode, the resulting effective IFE field would be inhomogeneous both in lateral and normal to the film directions which might result in the excitation of the standing spin waves~\cite{krichevsky2021selective,krichevsky2024spatially}. 

\begin{figure}[ht]
    \centering
\textbf{a}~~~~~~~~~~~~~~~~~~~~~~~~~~~~~~~~~~~~~~\textbf{b}~~~~~~~~~~~~~~~~~~~~~~~~~~~~~~~~~~~~\textbf{c}\\
    \includegraphics[width=0.32\linewidth]{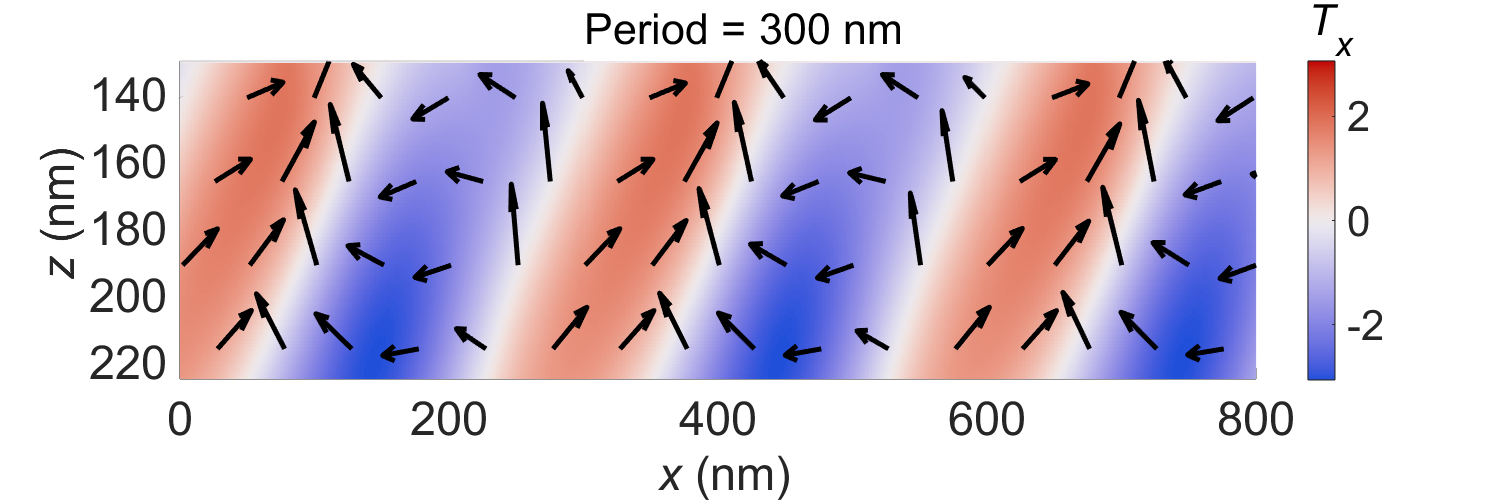}
    \includegraphics[width=0.32\linewidth]{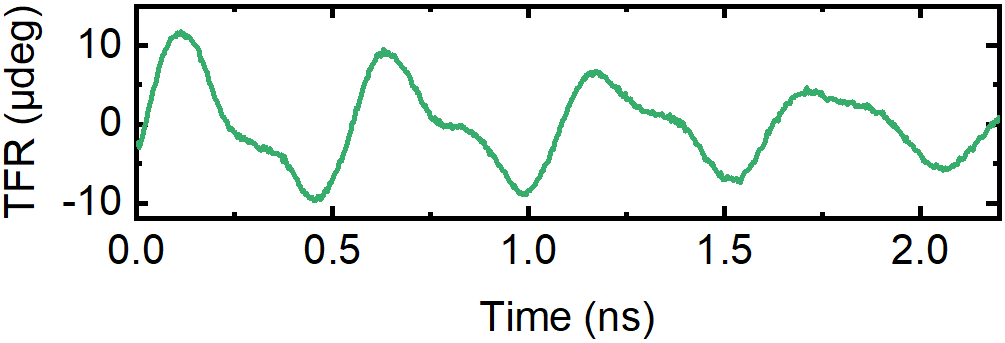}
    \includegraphics[width=0.32\linewidth] {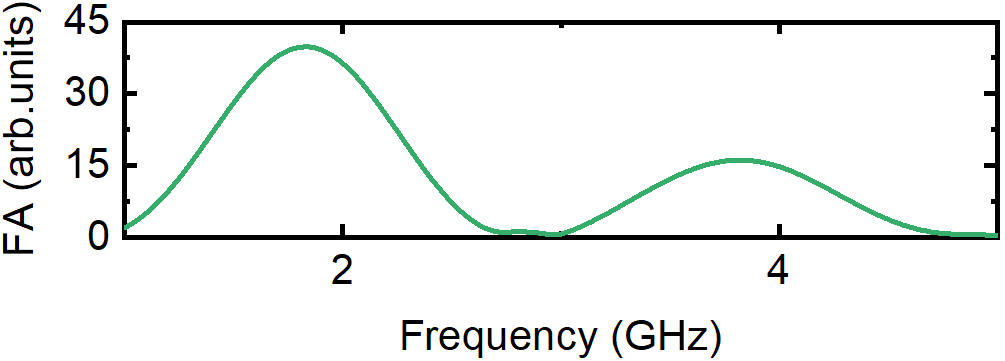} \\
 \textbf{d}~~~~~~~~~~~~~~~~~~~~~~~~~~~~~~~~~~~~~~\textbf{e}~~~~~~~~~~~~~~~~~~~~~~~~~~~~~~~~~~~~\textbf{f}\\
    \includegraphics[width=0.32\linewidth]{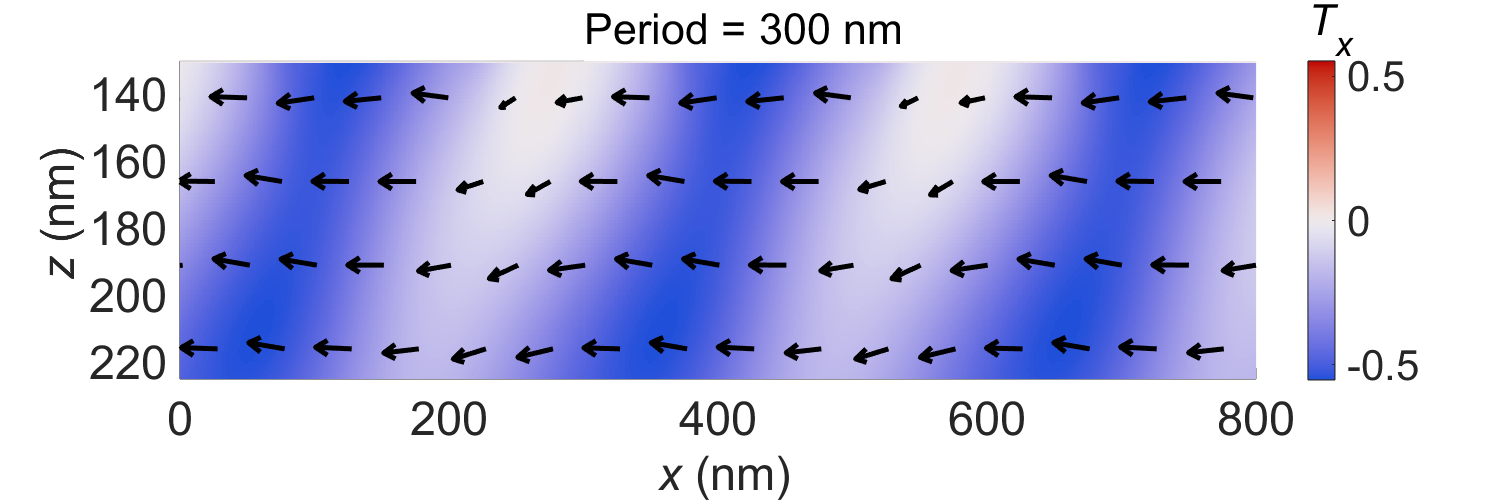}
    \includegraphics[width=0.32\linewidth]{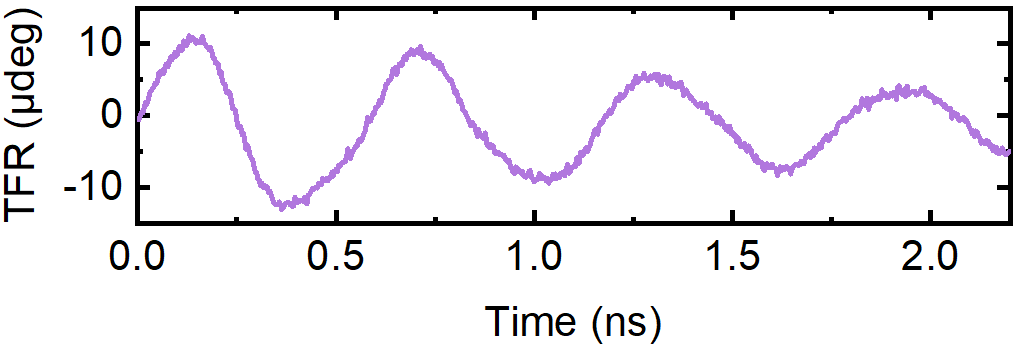}
    \includegraphics[width=0.32\linewidth]{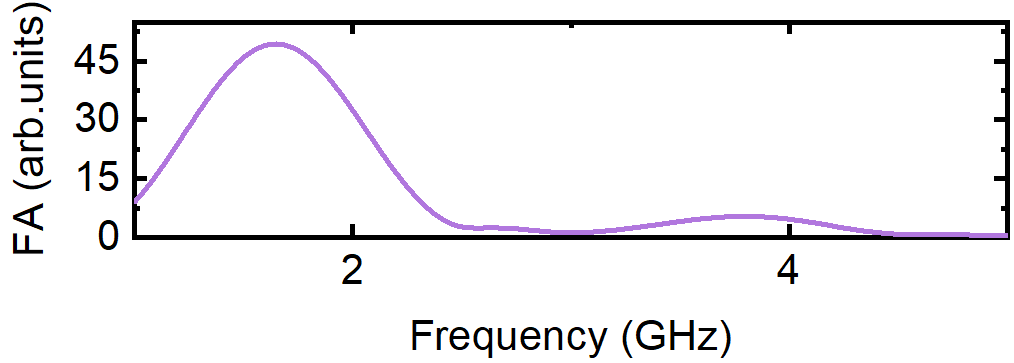}\\
 \textbf{g}~~~~~~~~~~~~~~~~~~~~~~~~~~~~~~~~~~~~~~\textbf{h}~~~~~~~~~~~~~~~~~~~~~~~~~~~~~~~~~~~~\textbf{i}\\
    \includegraphics[width=0.32\linewidth]{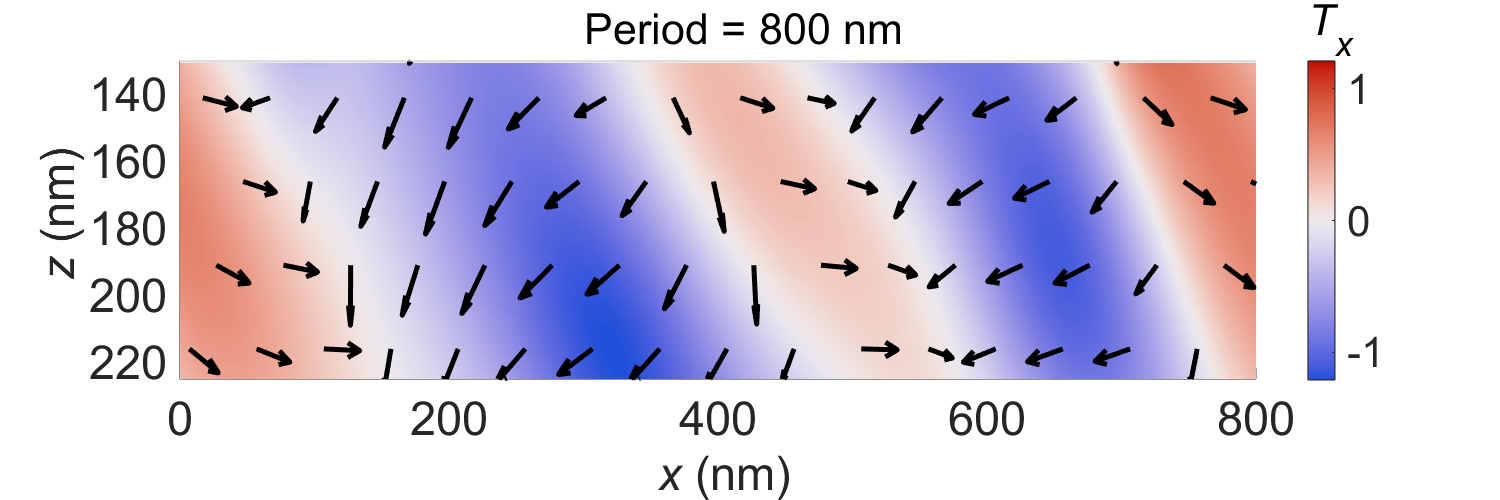}
    \includegraphics[width=0.32\linewidth]{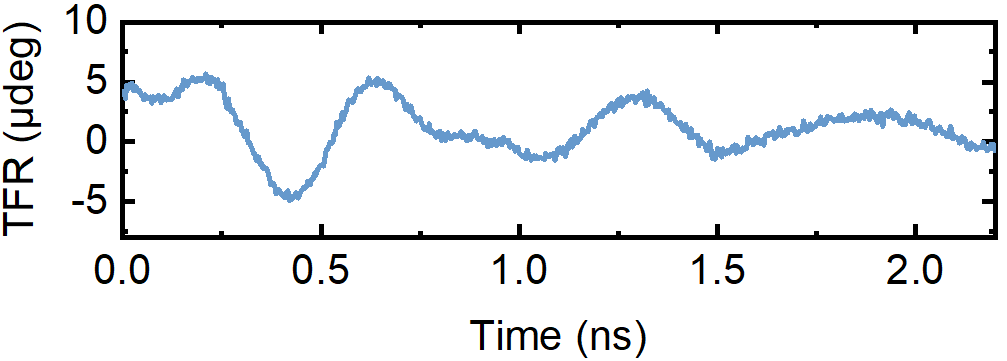}
    \includegraphics[width=0.32\linewidth]{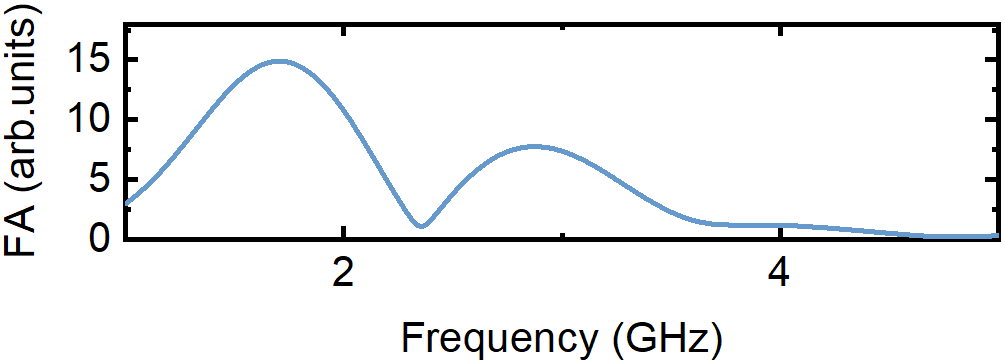}\\
  
    \caption{Spin dynamics induced in MPWG with $P=300$~nm (\textbf{a-f}) and $P=800$~nm (\textbf{g-i}) periods. \textbf{a,d,g} Distribution of $T_x$. The arrows indicate the direction of the torque. Only BIG part of one period of MPWG is shown for visibility. \textbf{b,c} and \textbf{h,i} Dynamic signals and their Fourier spectra at the resonant wavelengths of the studied gratings, 685~nm (\textbf{b,c,h,i}) and 715~nm (\textbf{e,f}).}
    \label{Fig: Precession diff cases}
\end{figure}

The spatial period of the IFE torque induced by the TE-mode, $P_t$ is determined by its dispersion and diffraction order $m$ that excites the guided mode at a given wavelength. Thus, period of $\textbf{T}(\textbf{r})$ equals to $P_t=P/m$ and may coincide with the period of the grating $P$ or be less by an integer number of times $m$. It is important to underline that this is a feature only of the resonantly pumped MPWG. If the pump pulse wavelength is detuned from its resonance, for example, to 750~nm (Fig.~\ref{Fig: Optical properties}a, black circle), the IFE torque becomes nearly uniform (compare Fig.~\ref{Fig: Precession diff cases}a and d) except for some inhomogeneity brought by the non-resonant light scattering on a grating which is qualitatively quite similar to the case of the bare film. As a result, shifting pump wavelength from the resonant $\lambda=685~nm$ to a non-resonant $\lambda=750~nm$ drastically changes character of the observed signal: the precession becomes very close to the decaying harmonics (Fig.~\ref{Fig: Precession diff cases}e) and the high-frequency peak nearly vanishes (Fig.~\ref{Fig: Precession diff cases}f).

To demonstrate an ability to tune the wavelength of the excited spin waves, the MPWG structure with period $P=800$~nm and 610-nm wide TiO$_2$ stripes was fabricated. It supports excitation of the TE guided mode at $\lambda=715~nm$ and $\theta=17^\circ$ angle of incidence (see Methods for the detailed information on its optical properties). This excitation is performed by $m=2$ grating diffraction order, so that the phase of $E_y$ component changes for $4\pi$ on the scale of one period. Consequently, $\textbf{T}(\textbf{r})$ is also sign-changing and periodic with a period equal to $P/m=400$~nm (Fig.~\ref{Fig: Precession diff cases}g).

Figure~\ref{Fig: Precession diff cases} summarizes the excited spin dynamics and its spectra for the three cases discussed above: resonant and non-resonant wavelengths of 300-nm-period waveguide grating and resonant wavelength of 800-nm-period waveguide grating. One might see a noticeable difference in the observed signals (Fig.~\ref{Fig: Precession diff cases}b,e,h). All of the waveguide gratings produce the same low-frequency peak at $f=1.9$~GHz corresponding to the long-wavelength spin wave with $K\approx 0$. The high-frequency peaks are observed only for the two resonantly pumped waveguide gratings (Fig.~\ref{Fig: Precession diff cases}c,i). These peaks have different positions, $f=3.9$~GHz and $f=2.9$~GHz for  300-nm and 800-nm period MPWG structures, respectively. 

\section*{Optical excitation of the short spin waves in the magnetophotonic waveguide grating}\label{sec5}

\begin{figure}[htb]
    \centering ~~~\textbf{a}~~~~~~~~~~~~~~~~~~~~~~~~~~~~~~~~~~~~~~~~~~~~~\textbf{b}\\
        \includegraphics[width=0.4\linewidth]{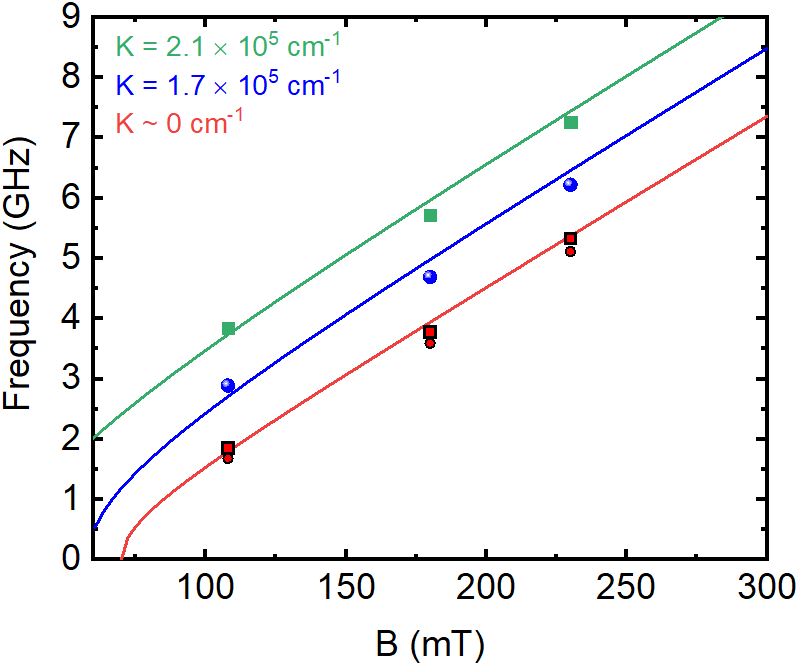}
        \includegraphics[width=0.4\linewidth]{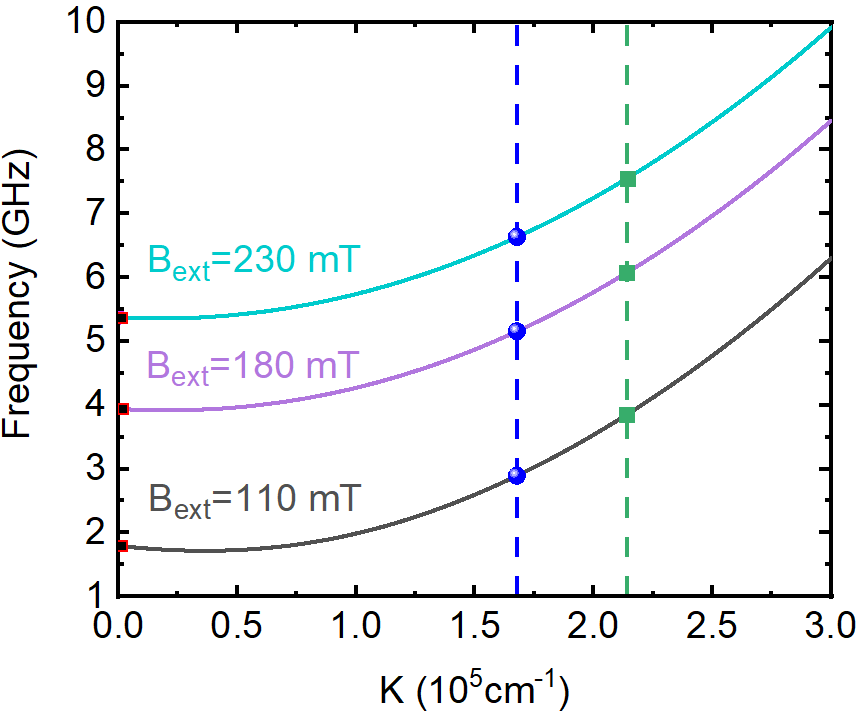}

    \caption{Dispersion of the spin waves excited in MPWG. Solid curves represent calculated data while color symbols show experimentally found data for the grating of period $P=800$~nm (red triangles for the low frequency and blue circles for the high frequency modes), and grating of period $P=300$~nm (red squares for the low frequency and green squares for the high frequency modes).  \textbf{a,}  Dependence of the spin-wave frequency  on the external magnetic field $B_\mathrm{ext}$. Solid curves represent calculations performed using Eq.~\eqref{Eq SW frequency}  for the spin-wave wavenumbers $K\approx 0$, $K=1.6\cdot 10^{5}~\mathrm{cm}^{-1}$ and $K=2.1\cdot 10^{5}~\mathrm{cm}^{-1}$ (red, blue and green colors, respectively). \textbf{b,} Dispersion of spin-waves $f(K)$ for $B_\mathrm{ext}=110,180,230$~mT (black, violet and light-blue colors, respectively).}
    \label{Fig: SW frequencies}
\end{figure}

The periodic pattern of the inverse Faraday effect and its torque acting on magnetization excite not only spin modes with vanishing wavenumbers but also modes with much higher wavenumbers. Indeed, let’s approximate the distribution of the in-plane torque by $T_x(\mathbf r) = T_0 \exp(-\frac{r^2}{r_0^2})(1+2\cdot \sin(2\pi x/P_t))$. Then its Fourier transform is given by

\begin{eqnarray}
T_x(\mathbf K)= \frac{T_0 r_0^2}{4} \Biggl[2 \exp\Bigl(-\frac{r_0^2}{4} K_y^2\Bigr)\biggl(\exp\Bigl(-\frac{r_0^2}{4}(K_x+\frac{2\pi}{P_t})^2\Bigr) - \exp\Bigl(-\frac{r_0^2}{4}(K_x-\frac{2\pi}{P_t})^2\Bigr)\biggr) -\nonumber \\ 
-i\exp\Bigl(-\frac{r_0^2}{4} (\mathbf K ^2)\Bigr)\Biggr].
\end{eqnarray}

Consequently, it can launch spin waves with wavenumbers along x-axis as large as $K_x=2\pi/P_t$. In the case of the experimentally studied gratings $K_x=2.1\cdot 10^{5}~\mathrm{cm}^{-1}$ and $K_y=1.6\cdot 10^{5}~\mathrm{cm}^{-1}$ corresponding to the gratings with $P=300$~nm and $P=800$~nm, respectively. The exchange interaction becomes important for such short spin waves. At the same time, the thickness of the film is comparable to the wavelength, so the magnetodipole interaction should also be taken into account. In addition, the considered film possesses uniaxial magnetic anisotropy, while cubic anisotropy can be neglected. Pinning parameters can be assumed to be zero. Thus, the model derived by Kalinikos and Slavin was chosen for the theoretical description of spin waves $f(K)$ dispersion~\cite{kalinikos1990dipole} for the propagation along x-axis:
\begin{equation}
f(K_x) = \gamma \sqrt{\left(B_\mathrm{ext}+\frac{2A}{M_s}K_x^2+\mu_0 M_s\frac{  K_x  d}{2}\right)\left(B_\mathrm{ext}+\mu_0 M_s (1- \frac{K_x  d}{2})-\mu_0 H_u+\frac{2A}{M_s}K_x^2\right)}, \label{Eq SW frequency}
\end{equation}
where $M_s = 4.0\cdot10^4$~A/m is a saturation magnetization, and $H_u = 2K_u/{\mu_0 M_s}$ is a uniaxial anisotropy field, $A$ is an exchange stiffness constant. For the studied film $K_u$ = 2390 N/m$^2$ and $A = 4$~pJ/m.  

Figure~\ref{Fig: SW frequencies}a demonstrates a good agreement of theoretically estimated spin wave frequencies with experimental results obtained for the different external magnetic fields for both of the waveguide gratings. Therefore, the observed high frequency peak of the excited spin dynamics (see Fig.~\ref{Fig: Precession diff cases}) is explained by excitation of short wavelength spin waves. The dispersion in Fig.~\ref{Fig: SW frequencies}b clearly shows that the waveguide gratings excite the narrow band of short spin waves. The narrow band is provided by the periodic subdiffractive distribution the IFE torque which results in a significant increase of the lifetime of the observed spin dynamics (compare Fig.~\ref{Fig: Scheme and precession}b and d) from 0.8 ns for the bare magnetic film to 3.8 ns for the magnetic film covered by a dielectric grating. 

The spin waves have a character of the backward volume type ($V_\mathrm{gr}<0$) for relatively small wavenumbers (for $K< 0.3\cdot 10^5~\mathrm{cm}^{-1}$) and become of the forward kind ($V_\mathrm{gr}>0$) in the region where the exchange interaction prevails for ($K> 0.3\cdot 10^5~\mathrm{cm}^{-1}$). Notice that despite the configuration refers to the 'backward' volume spin waves excitation, actually the excited short spin waves are forward ($V_\mathrm{gr}>0$) due to the high exchange contribution at such spatial scale.

Such waves are interesting for practical applications from several points of view. First of all, submicron wavelengths make them promising for miniaturizing optomagnonic spin-wave devices. On the other hand, the frequencies of such waves are larger due to the exchange contribution, and allow for the higher performance of such optomagnonic devices. Finally, spin-waves with $K\sim 2 \cdot 10^5~\mathrm{cm}^{-1}$ have higher group velocities determined by $2\pi\cdot \partial f / \partial K$ that make them faster, and smaller group velocity dispersion  $2\pi\cdot \partial^2 f / \partial K^2$ that reduces the spin pulse dispersion spreading. Namely, the group velocity of the excited spin wave with $K= 2.1 \cdot 10^5~\mathrm{cm}^{-1}$ is $V_\mathrm{gr}=1.38\cdot 10^3~\mathrm{m\cdot s}^{-1}$ for the external magnetic field $B_\mathrm{ext}=180$~mT. It is an order higher than the group velocity for $K=0$ where $V_\mathrm{gr}=-0.14\cdot 10^3~\mathrm{m\cdot s}^{-1}$. Such short waves are faster than the domain wall (750 m/s ~\cite{yang2015domain}) and skyrmion motions (900 m/s~\cite{pham2024fast}). Further modification of the iron garnet composition, namely the reduction of the saturation magnetization, might increase the group velocity even more (see Fig.~\ref{Fig: Discussion}b).

\section*{Optical detection of short spin waves}

 \begin{figure}[htb]
     \centering
\textbf{a}~~~~~~~~~~~~~~~~~~~~~~~~~~~~~~~~~~~\textbf{b}~~~~~~~~~~~~~~~~~~~~~~~~~~~~~~~~~~~\textbf{c}\\~\\
    \includegraphics[width=0.32\linewidth]{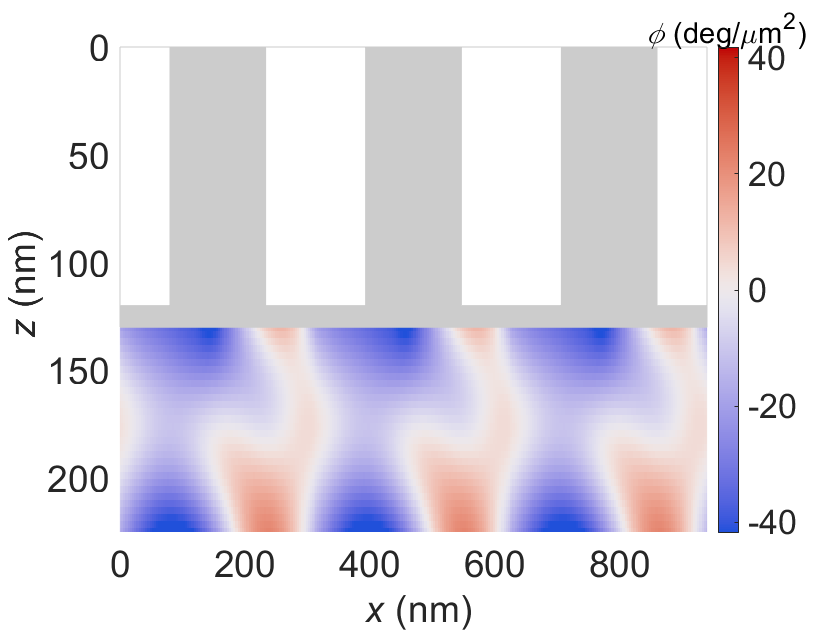}
    \includegraphics[width=0.32\linewidth]{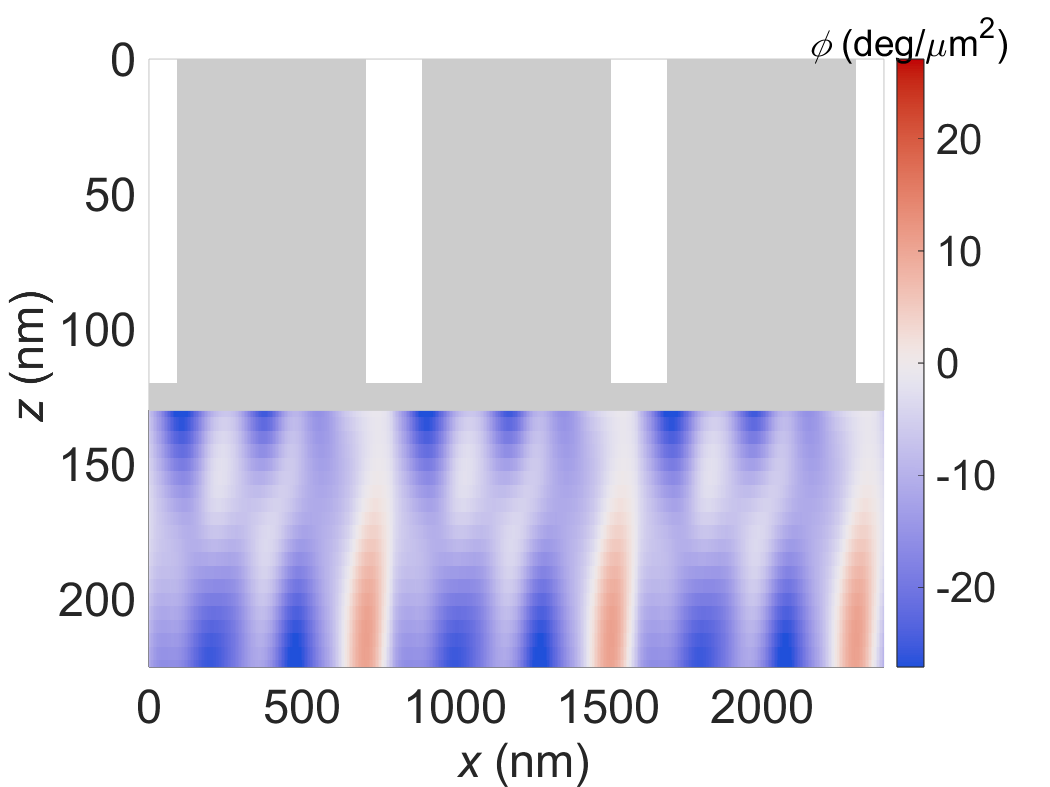}
    \includegraphics[width=0.32\linewidth]{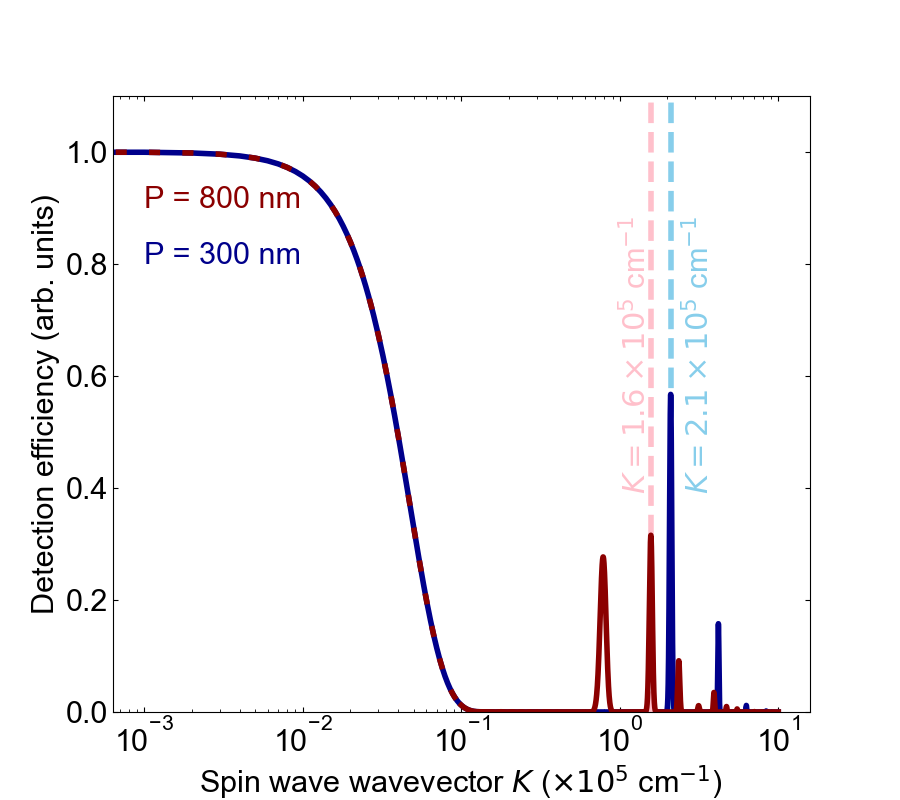}\\
    \caption{\textbf{Detection of submicron spin waves in the MPWG.} \textbf{a,b,} The Faraday rotation contribution $\phi$ of each part of the magnetic BIG film of the MPWG with periods $P=300$~nm (\textbf{a,}) and $P=800$~nm (\textbf{b,}) calculated for a 525-nm probe pulse of 7~$\mu$m diameter used in our experiments. \textbf{c,} Efficiency of the magneto-optical spin-wave detection in MPWG with periods $P=300$~nm (blue curve) and $P=800$~nm (red curve) by the probe pulse with 525~nm wavelength and 7~$\mu$m diameter. Vertical dashed lines note wavenumbers of spin waves excited by the optical pump.}
    \label{Fig: Optical detection}
\end{figure}

Optical detection of subdiffractive spin wave is not less challenging than their excitation since the principle of spin dynamics measurements in a pump-probe setup is based on the measurement of the probe pulse Faraday rotation averaged over the beam area. This makes the detection of the short spin waves  with a wavelength smaller than the probe beam size impossible in the smooth films since the average magnetization in the region covered by a probe beam is zero~\cite{ignatyeva2022magneto}. Similarly to the case of pump beam, the diffraction limit also imposes the restrictions on the ability to focus the probe, so usually it is several micrometers in radius or wider. Thus, short spin waves with the submicron wavelength discussed above are unreachable for detection in ordinary setups.

 However, the situation changes if the smooth bare film is replaced by a nanophotonic structure. The inhomogeneity of the light distribution in the nanostructure makes it inhomogeneously sensitive to the magnetization of the different regions of a magnetic material~\cite{ignatyeva2022magneto}. Namely, the periodicity of the distribution of the probe electromagnetic field provided by a grating allows to detect the spin waves with a wavelength equal or multiple times smaller than the grating period~\cite{borovkova2022spectrally,ignatyeva2022magneto}. Figures~\ref{Fig: Optical detection}a,b show the specific contribution to the Faraday rotation of each part of the magnetic BIG film under $P=300$~nm and $P=800$~nm gratings calculated for a probe pulse with a wavelength 525~nm and 3.5~$\mu$m radius used in our experiments. It's sign-changing and periodic. The period of the Faraday rotation contribution function in $P=300$~nm waveguide grating (Fig.~\ref{Fig: Optical detection}a) coincides with a grating period. The pattern of the Faraday contribution function in $P=800$~nm waveguide grating (Fig.~\ref{Fig: Optical detection}b) is more complicated and obviously contains several $P/m$ harmonics. Since the sensitivity of the Faraday effect depends on the lattice period, it is possible to measure the dynamics of the spin waves with wavelengths less than the probe pulse diameter.  
 
 Figure~\ref{Fig: Optical detection}c shows the theoretical estimations of the magneto-optical probing efficiency calculated for the different spin-wave wavelengths. One might see that the optical probe in the magneto-optical setup is sensitive both to the long spin-wave wavelengths ($K<10^{4}\mathrm{cm}^{-1}$) and to the short spin waves with the wavelengths $\Lambda=P$ ($K=2.1\cdot10^{5}\mathrm{cm}^{-1}$). The detection efficiency of these low and high-K spin waves is similar, so we are able to see both peaks in the spectra of the probe Faraday rotation caused by the magnetization precession during the spin-wave propagation.

\section*{Outlook}

 \begin{figure}[htb]
     \centering
\textbf{a}~~~~~~~~~~~~~~~~~~~~~~~~~~~~~~~~~~~~\textbf{b}\\
    \includegraphics[width=0.32\linewidth]{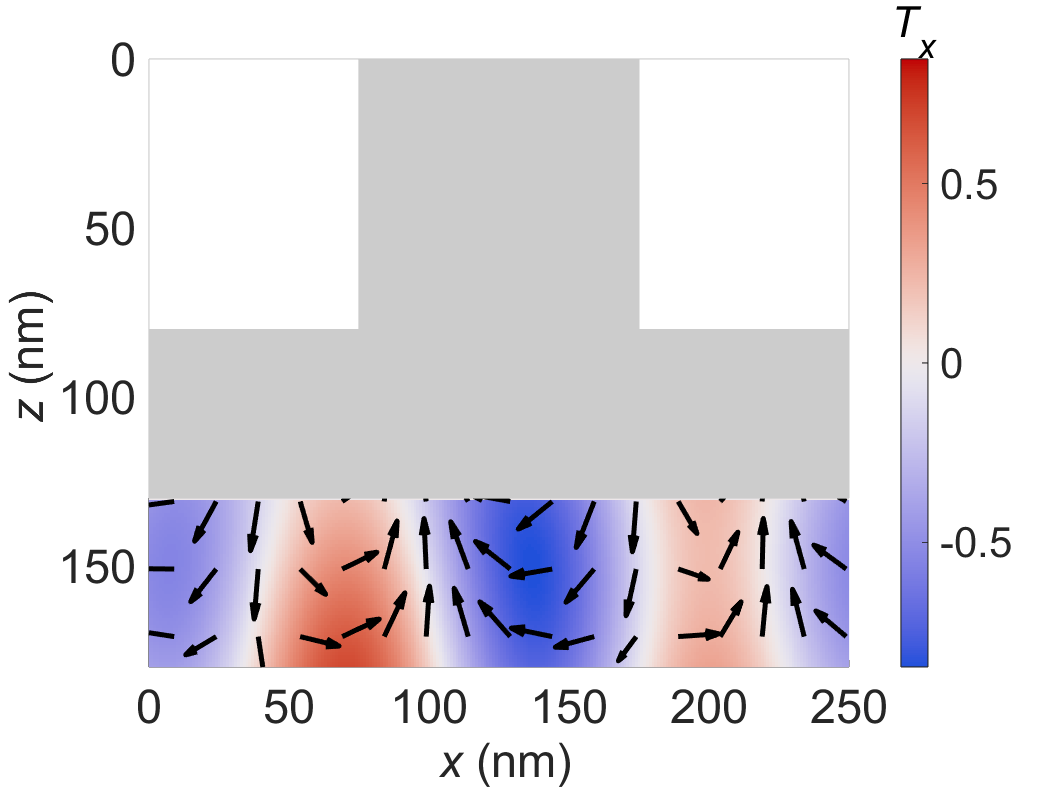}
     \includegraphics[width=0.32\linewidth]{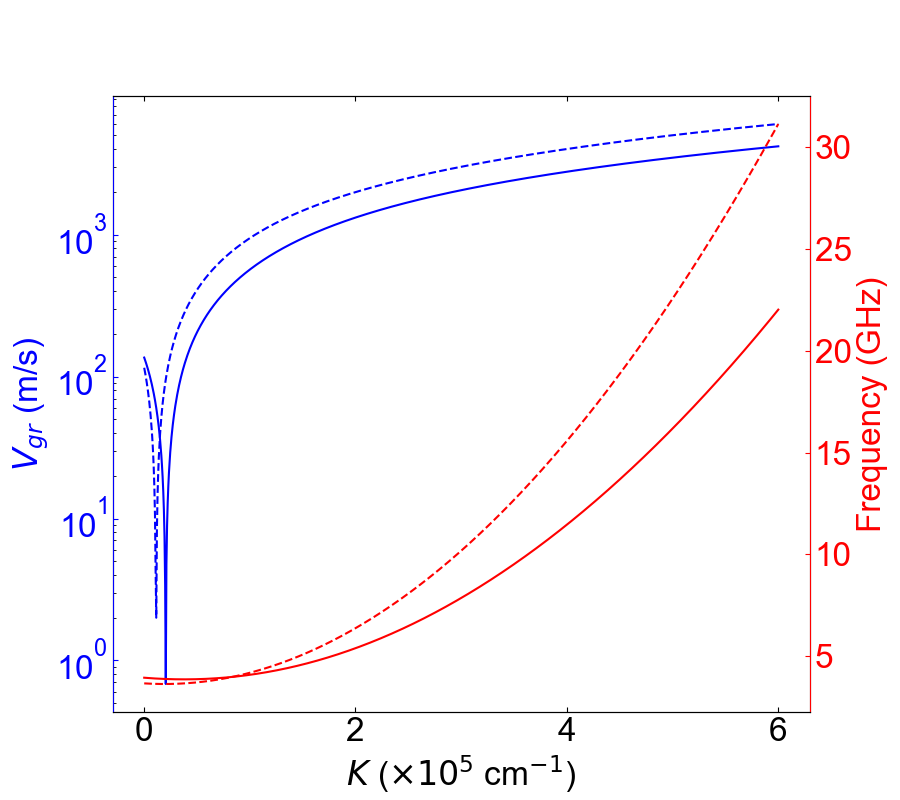}
    \\
    \caption{\textbf{Perspective of further short spin-wave device improvement.} \textbf{a,} $T_x$ in the MPWG with Si and BIG layers (see parameters in the text) providing IFE distribution for excitation of the spin waves with $\Lambda=125$~nm ($K=5\cdot10^5~\mathrm{cm}^{-1}$ ).  \textbf{b,} Group velocity (blue color) and spin-wave frequency (red color) depending on the spin-wave wavevector K. Solid lines represent the calculations performed for the studied BIG film. Dashed lines correspond to the BIG film with saturation magnetization equal to $2.8\cdot10^4$~A/m.}
    \label{Fig: Discussion}
\end{figure}

The MPWG stucture in which the TE-mode is excited creates the inhomogeneous distribution of the inverse Faraday effect and allows for the excitation of the short spin waves with a wavelength coinciding with the grating period or integer times smaller it. It is important to underline that the physical origin of such a sign-changing inverse Faraday effect behaviour at the scale of the grating period is in the superposition of the TE-mode propagating along the waveguide grating periodicity direction and non-resonant quasi-uniform TM components of light. Thus, the presence of the optical waveguide resonance is essential for the short spin wave excitation so that the period of the induced IFE field cannot be scaled arbitrary.

Thus, there are two issues that are important for further applications of the proposed MPWG structures. The first one is the smallest wavelengths that can be excited using the proposed structure. The second one is an ability to tune the parameters of the excited spin waves.

Firstly, we focus on how short spin wave can be optically excited. As it was discussed above, the spin wave wavelength is given by $\Lambda=P_t=P/m$. On the other hand, the proposed mechanism implies that a waveguide mode is excited. It follows from Equation~\eqref{Eq: WG disp}, that minimal spin wave wavelength can be found as $\Lambda_\mathrm{min}=\lambda/(n_\mathrm{wg} \pm \sin\theta)$. Consequently, one should minimize the pump’s wavelength and maximize the refractive index $n_\mathrm{wg}$ of the guided mode. At the same time, non-thermal excitation of spin dynamics in iron-garnets by IFE could be implemented only in its transparency window. This limits possible wavelengths for iron garnet films to $\lambda\gtrapprox550$~nm. Secondly, one might use an oblique incidence with rather large $\theta$. Finally, high refractive index of the waveguide mode $n_\mathrm{wg}$ determined by the design of MPWG can be made larger, up to $n_\mathrm{wg}\approx 4.5$ if MPWG with high-refractive index semiconductors, such as Si, GaP, etc., is used. It is important to emphasize here that using such nonmagnetic part for the hybrid MPWG allows to excite a guided mode whose refractive index is about two times larger than for the iron garnet film (for the iron garnet $n \approx 2.3$). In this case the guided mode is mainly localized in the nonmagnetic part of MPWG but still influences on the magnetic film below by its near field. Consequently, the above reasoning estimates the shortest spin wave non-thermally excited by light in transparent garnets as $\Lambda_\mathrm{min} \approx \lambda/5.5 \approx 100$~nm.

To confirm this, we performed numerical simulations of the IFE induced torque $T_x(\textbf{r})$ distribution in a hybrid Si-BIG MPWG structure with the following parameters: $P=250$~nm,  thickness of the BIG $d=50$~nm and Si layers $d_\mathrm{Si}=50+80$~nm for the smooth+etched parts, correspondingly. The waveguide structure was pumped by the circularly polarized light with $\lambda=558$~nm and $\theta=17^\circ$, see Fig.~\ref{Fig: Discussion}. The IFE distribution period is 2 times smaller than the grating period and corresponds to the spin wave wavelength of $\Lambda=125$~nm or $K=5\cdot10^5~\mathrm{cm}^{-1}$.

As it was mentioned above, the deposition of the waveguide grating fixes the period of the IFE distribution as $P/m$ value. Thus, tuning of the launched short spin wave wavelength could be performed by the excitation of the waveguide resonances with different $m$ at different light wavelengths or angles of incidence. Such an approach allows to excite a discrete set of the spin-wave wavelengths at a fixed spatial position. On the other hand, one might design a 'chirped' waveguide grating with the period changing along the stripe. This might allow to achieve a continuous change of the excited spin-wave wavelength along the grating stripe. At the same time, both of these approaches in some sense fix the position and the wavelengths of the spin waves. It is possible to further change the configuration in order to get even more tunable system if the fabricated gratings are replaced with the optically induced ones.

Two linearly polarized light beams with orthogonal polarization directions produce the inverse Faraday effect pattern $H_z^\mathrm{IFE}\propto \mathrm{Im}(E_xE^*_y - E_y E^*_x)$. For the beams with the orthogonal polarization and opposite incidence angles $\pm \theta$ complex electric fields in the BIG film with $n_\mathrm{BIG}$ refractive index can be written as $E_{x,y}=E_0 \exp(\pm i k_0 n_\mathrm{BIG} \sin(\theta) x)$. Thus IFE pattern created by such two-component field is periodic and sign-changing: $H_z^\mathrm{IFE}\propto E_0^2 \sin{\left(2 k_0 \sin\theta x\right)}$. Such distribution launches the short spin waves with a wavelength equal to $\Lambda=\lambda/(2 \sin\theta)$. This allows for a continuous change of the wavelength of spin wave optically excited in the iron garnet film from $\Lambda\approx 275$~nm to $\Lambda\rightarrow \infty$ via tuning of the incidence angles. 

\section*{Conclusion}\label{sec9}

A novel method for ultrafast optical excitation of subdiffractive spin waves with wavelengths of a few hundred of nanometers is demonstrated. It is based on covering of a dielectric magnetic film with a non-magnetic grating which makes the entire sample a kind of magnetophotonic waveguide grating supporting propagation of optical guided modes. The key point is to tune wavelength and incidence angle of the optical femtosecond pump beam to excite TE-mode which provides sign-changing periodic profile of the inverse Faraday effect and therefore optically induced torque acting on magnetization during the pulse duration. The spatial period of the IFE inside the magnetic film coincides with a period of a waveguide grating or is integer times smaller. It allows to launch short spin waves whose wavelength equals to the spatial period of the inverse Faraday effect distribution. 

Excitation of the short spin waves with the wavelengths 300~nm and 400 nm was experimentally demonstrated in the two different waveguide gratings. Importantly, MPWG provides excitation of the short spin waves in a very narrow band thus making the decay times of the observed spin dynamics much longer up to 10 ns. Moreover, such short spin waves are exchange dominated and have positive group velocity in the contrary to the spin waves excited in the same field configuration but in the bare magnetic film. Moreover, for the short spin waves group velocity is an order of magnitude higher than for the spin wave $K\approx 0$ which have magnetostatic character. 

These results are significant for miniaturization and acceleration of the optomagnonic devices with optically launched spin dynamics. Further elaboration of the proposed concept might allow to excite nonthermally even shorter spin waves whose wavelength is as small as $\sim100$~nm and to perform a continuous variation of the spin-wave wavelength by using a kind of chirped grating structures or using optically induced gratings of the inverse Faraday effect.

\section*{Methods}\label{sec6}
\subsection*{Waveguide grating fabrication}\label{subsec6}
Iron garnet film of (Bi,Y,Lu)$_3$(Fe,Ga)$_5$O$_1$$_2$ with thickness of 95 nm was grown on a gadolinium gallium garnet substrate with the crystallographic axis (001) orientation using a liquid phase epitaxy technology. Then, TiO$_2$ with a thickness of 130 nm was deposited onto the film using electron beam evaporation. After this, a mask of Cr and SiO$_2$ was deposited by electron beam evaporation onto the TiO$_2$. An electron resist was applied to this mask. Then, using electron lithography, the metasurface was exposed according to a given pattern, namely, the first grating had a pattern with a period of 300 nm and an air gap of 160 nm, and the second had a period of 800 nm and an air gap of 190 nm. After exposure, the electron resist was developed in a cooled developer. The next step was the step-by-step etching of the mask and silicon in a plasma-chemical etching setup. First, the SiO$_2$ layer was etched in a mixture of CHF$_3$ and O$_2$ gases. Then, using a SiO$_2$ mask, Cr layer was etched by oxygen plasma. At the next stage, TiO$_2$ was etched through the Cr mask to a depth of 110 nm using a mixture of SF$_6$ and Ar gases. Residues of the Cr mask were removed with liquid chemicals in a special chromium etchant, which does not affect TiO$_2$. 20-nm layer of smooth TiO$_2$ was left intentionally to ensure that the BIG film was not exposed to etching process. 

\begin{figure}[ht]
     \centering
\textbf{a}~~~~~~~~~~~~~~~~~~~~~~~~~~~~~~~~~~~~~~~~~~~~\textbf{b}\\
    \includegraphics[width=0.4\linewidth]{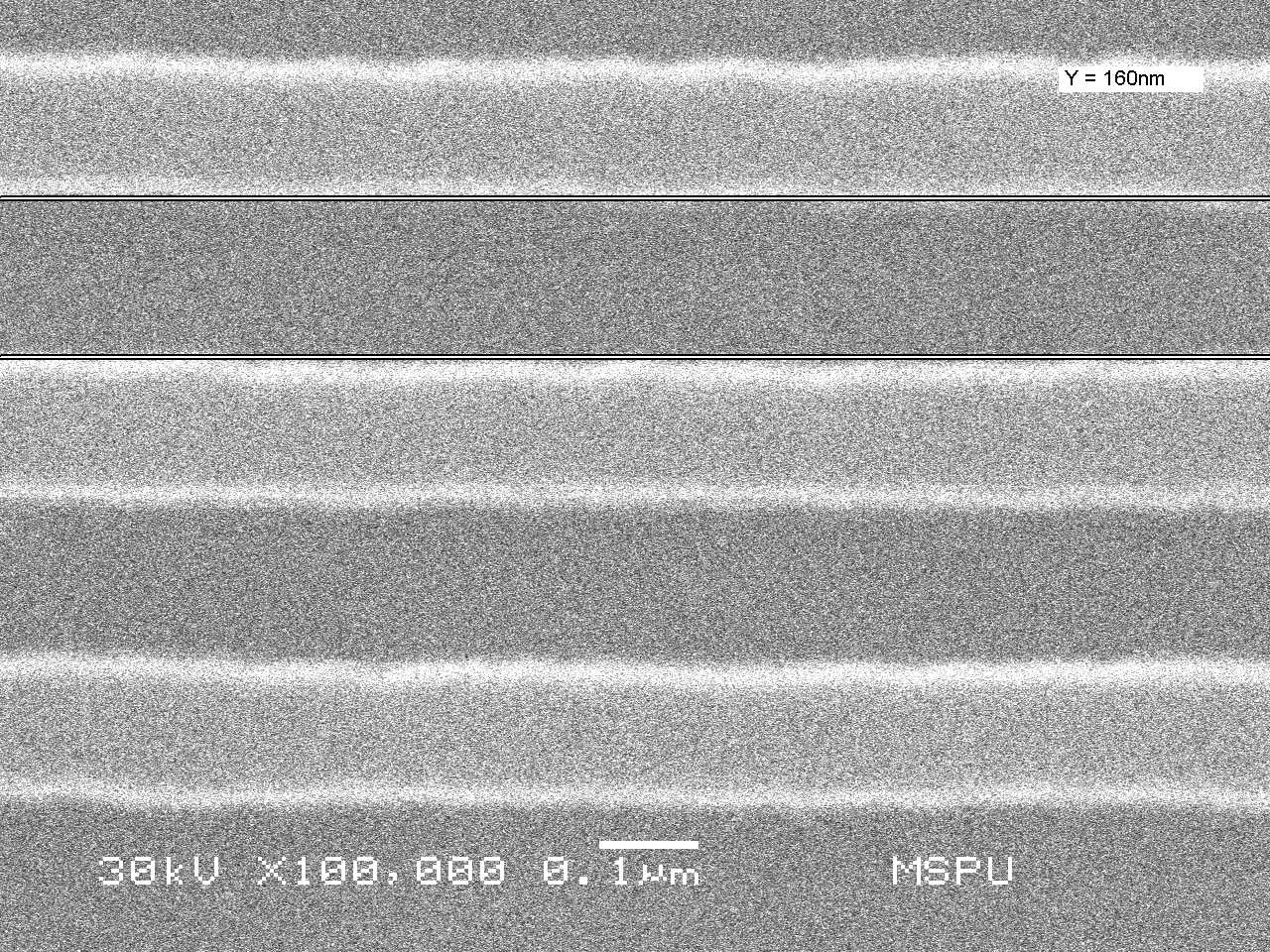}
    \includegraphics[width=0.4\linewidth]{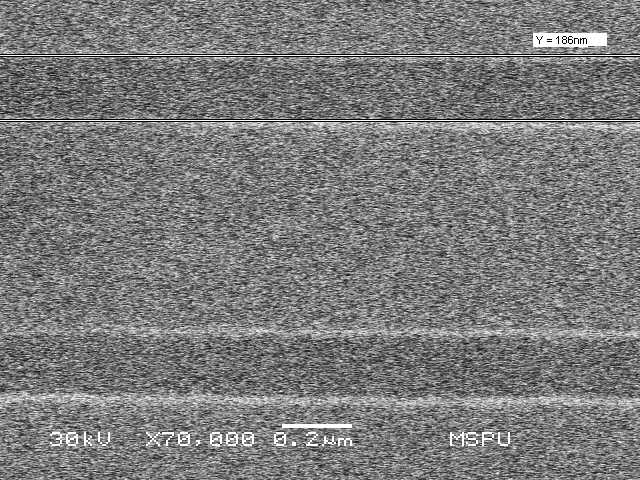}\\
    \caption{Scanning electron microscope images of the waveguide gratings with \textbf{a} ($P=300$~nm), \textbf{b} ($P=800$~nm) periods.}
    \label{Fig: SEM}
\end{figure}

Scanning electron miscroscope images shown in Fig.~\ref{Fig: SEM} demonstrate the resulting waveguide gratings nanopatterns. 

\subsection*{Experimental investigations}\label{subsec7}

First optical spectra of the waveguide gratings were measured to reveal the waveguide mode resonances using the following setup. A tungsten-halogen lamp was used as a light source. The light passed through the optical fiber. It was collimated by an achromatic 75-mm lens placed after the fiber, and then focused on the sample by another achromatic 35-mm lens into a spot with approximately 200~$\mu$m diameter. After light passed through the sample, the light was collimated by a microscope lens with 20x magnification and sent into the spectrometer slit. Thus the optical spectra were obtained simultaneously in a wavelength and angular domains.

\begin{figure}[ht]
     \centering  \textbf{a}~~~~~~~~~~~~~~~~~~~~~~~~~~~~~~~~~~~\textbf{b}~~~~~~~~~~~~~~~~~~~~~~~~~~~~~~~~~~~\textbf{c}\\
     \includegraphics[width=0.32\linewidth]{figures/Fig2.a45.png}
    \includegraphics[width=0.32\linewidth]{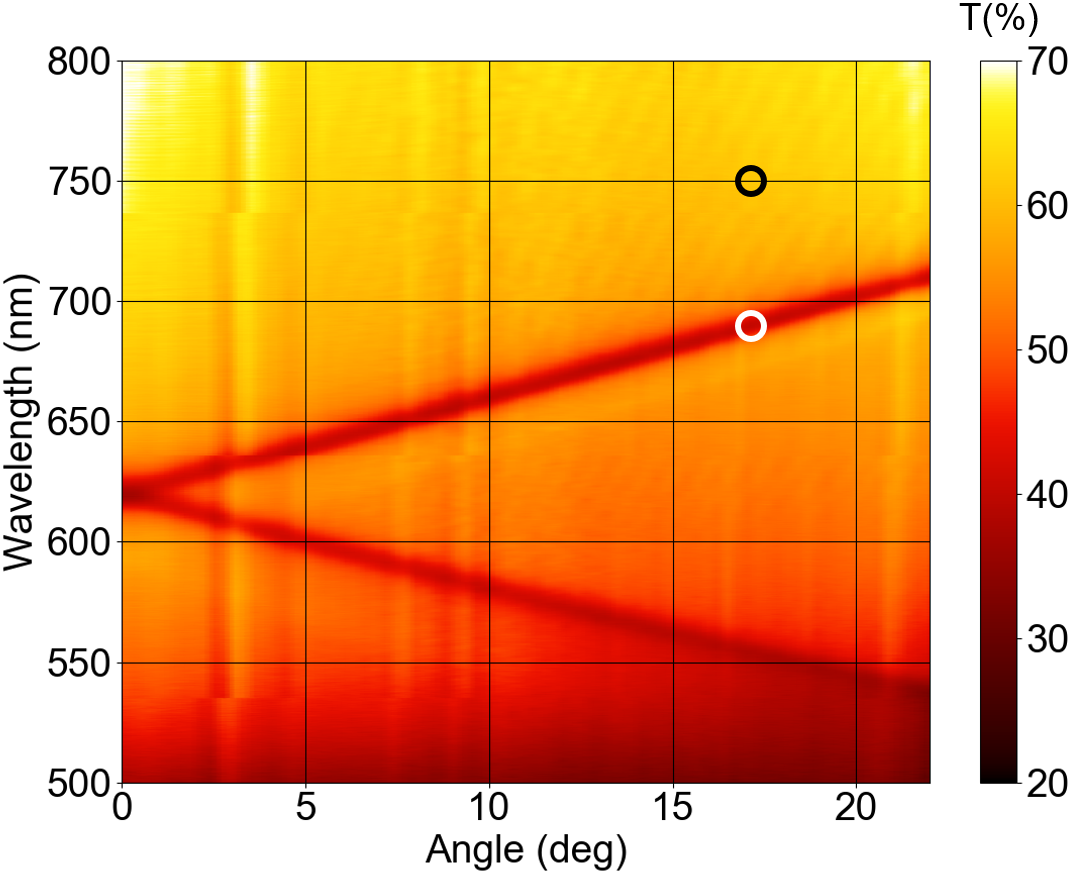}
     \includegraphics[width=0.32\linewidth]{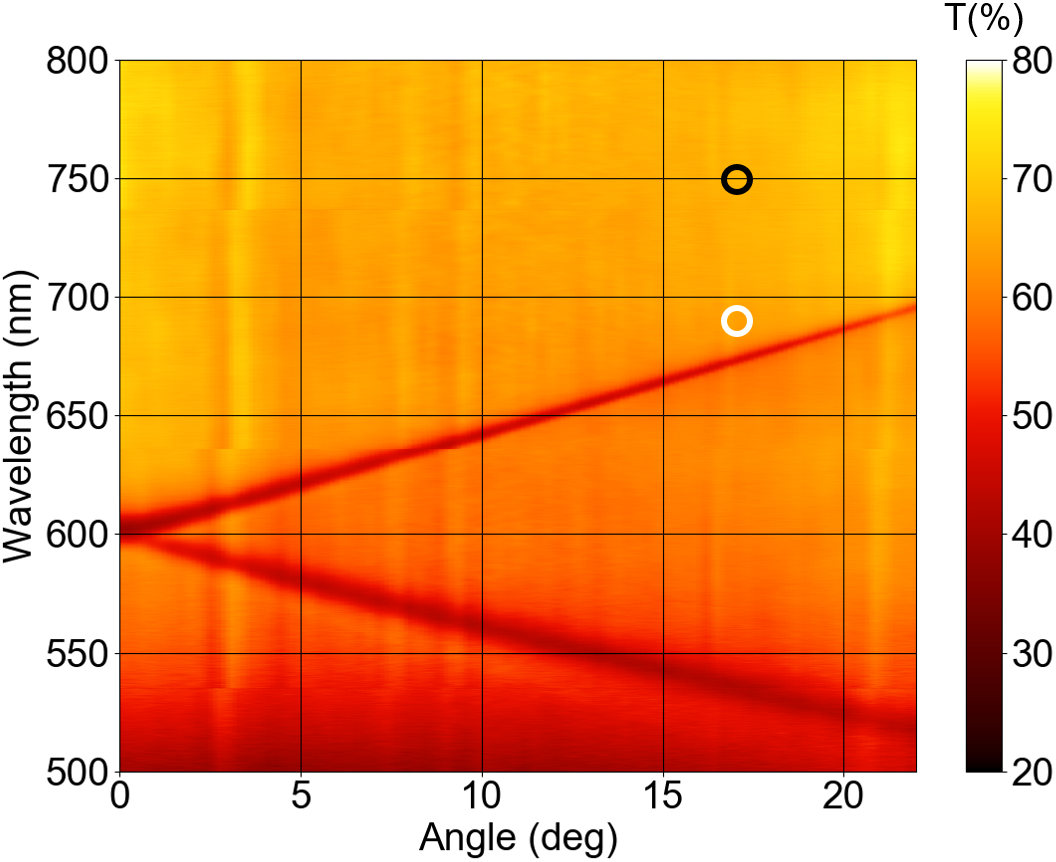}
 \textbf{d}~~~~~~~~~~~~~~~~~~~~~~~~~~~~~~~~~~~\textbf{e}~~~~~~~~~~~~~~~~~~~~~~~~~~~~~~~~~~~\textbf{f}\\
     \includegraphics[width=0.32\linewidth]{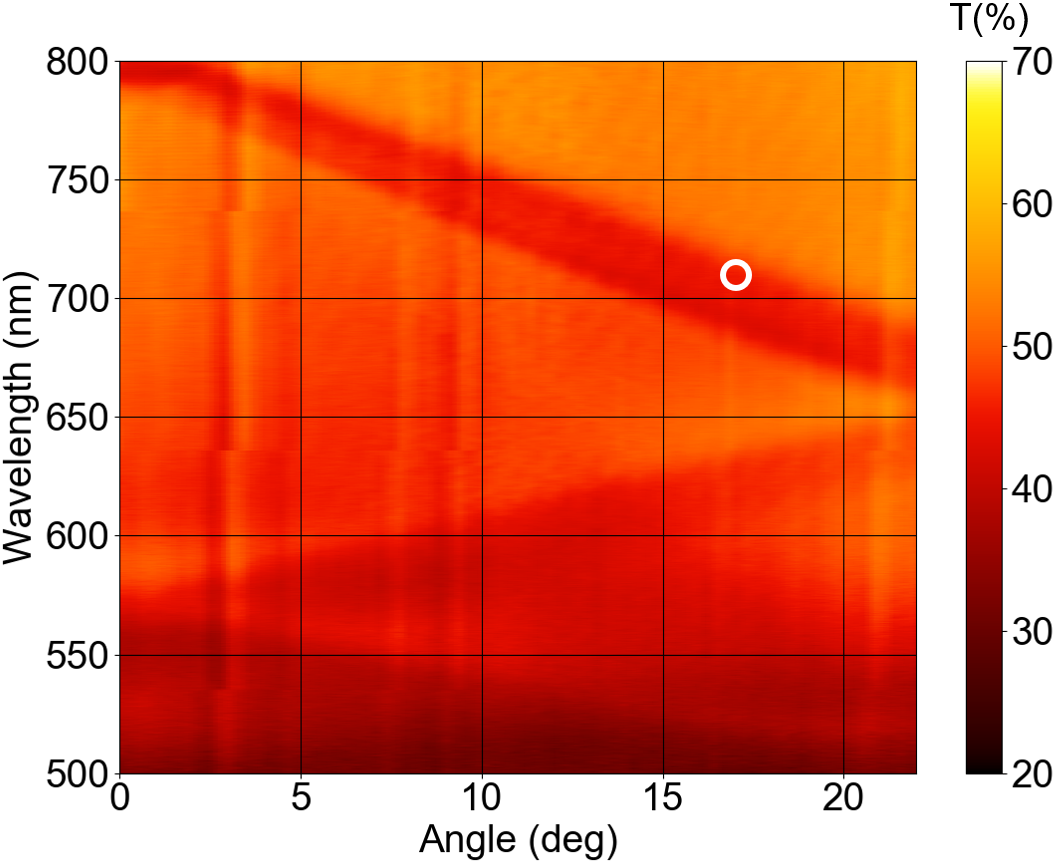}
    \includegraphics[width=0.32\linewidth]{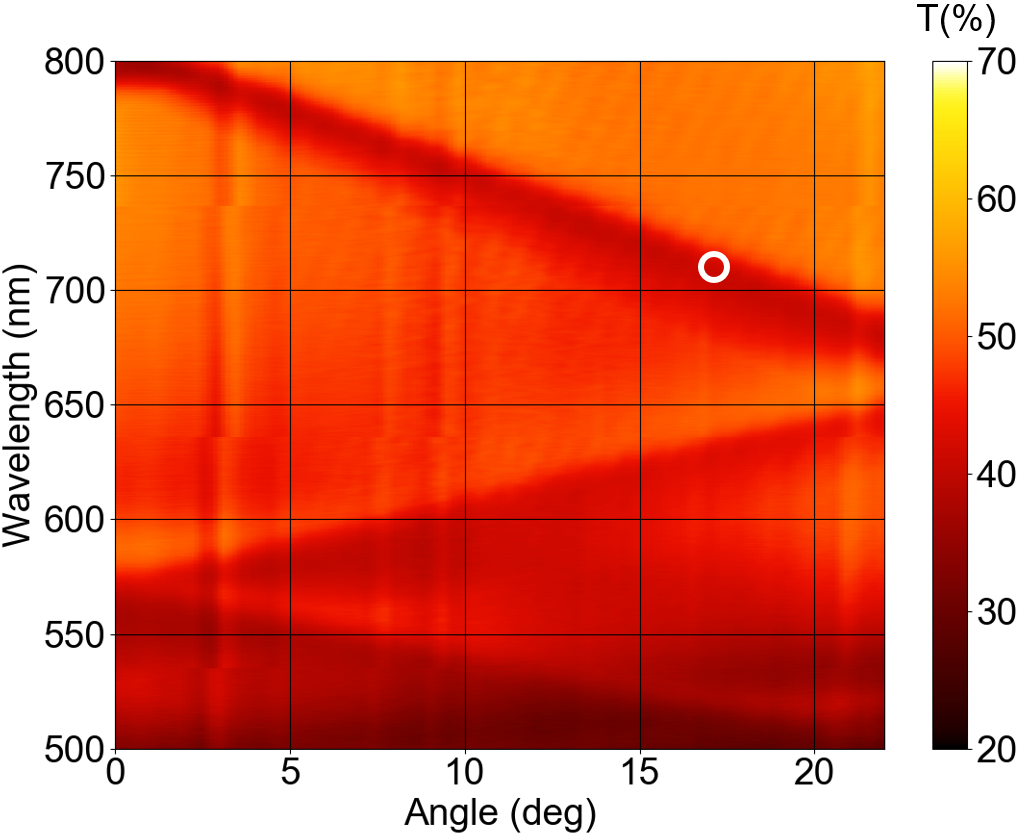}
        \includegraphics[width=0.32\linewidth]{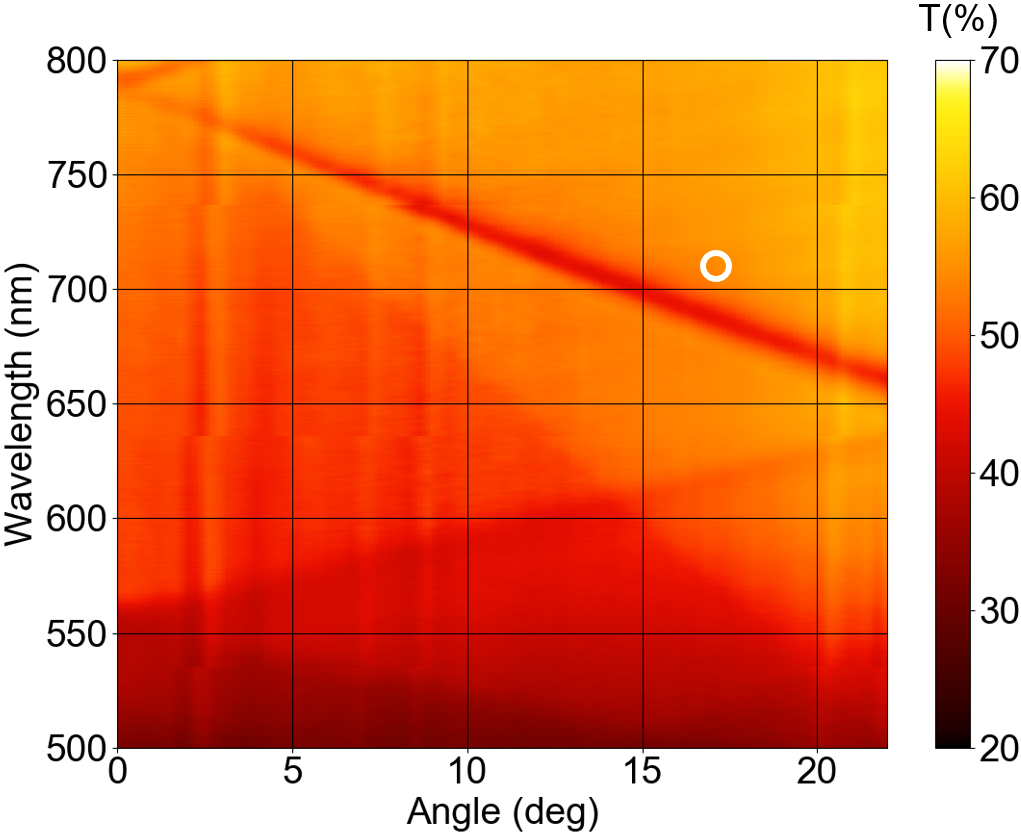}

    \caption{The experimental transmittance spectra of the waveguide gratings with 
 $P=300$~nm (top pane, \textbf{a-c}) and $P=800$~nm (bottom pane, top pane, \textbf{d-f}) measured for the circularly polarized (left column, top pane, \textbf{a,d}), s-polarized (central column, \textbf{b,e}) and p-polarized (right column, \textbf{c,f}) incident light. White circles indicate the TE-waveguide mode resonant wavelengths for the two gratings at $\theta=17^\circ$ angle of incidence. Black circle indicate the off-resonance wavelength used for a comparison purpose.}
    \label{Fig: Methods Grating spectra}
\end{figure}

The measurements were performed for the circular (Fig.~\ref{Fig: Methods Grating spectra}a,d), s-polarized (Fig.~\ref{Fig: Methods Grating spectra}b,e) and p-polarized (Fig.~\ref{Fig: Methods Grating spectra}c,f) light. It is well-known that the circularly polarized beam can be decomposed to the p- and s- polarized components with a $\pi/2$ phase shift between them. Thus, the presence of the certain resonance in the spectrum of p- or s-polarized light allows us to reveal the corresponding type of the mode excited in a waveguide grating under the illumination of the circularly polarized light. 

Spin dynamics experimental results were obtained using a time-resolved pump-probe setup. Pump and probe pulses were excited by a parametric oscillator Avesta TOPOL pumped by a Yb-dobed Avesta TEMA laser. This system allows for the generation of pulses with an 80 $\pm$ 5 MHz repetition rate and a 180 fs pulse duration. The wavelength of the probe was fixed and equal to 525 nm. The pump is tunable over a wide range of wavelengths (from 680 to 1050 nm) with spectral width about 6-8 nm. The diameter of the pump beam on the sample is approximately 7~$\mu$m. Three different wavelengths were used accorsing to the resonant ($\lambda=685$~nm for $P=300$~nm, $\lambda=715$~nm for $P=800$~nm), and non-resonant $750$~nm conditions. After passing through the delay line, pump pulses were modulated using a photoelastic modulator (Hinds Instruments PEM 100). Probe pulses with linear polarization were used to observe spin precession due to the magneto-optical Faraday effect. After the probe pulses passed through the sample, which was in an external magnetic field, an autobalanced optical receiver (Nirvana, 2007) used a synchronous detection scheme to record the change in probe pulse polarization.

\subsection*{Numerical simulations}\label{subsec8}

\begin{figure}[ht]
    \centering
     \centering  \textbf{a}~~~~~~~~~~~~~~~~~~~~~~~~~~~~~~~~~~\textbf{b}~~~~~~~~~~~~~~~~~~~~~~~~~~~~~~~~~\textbf{c}\\
    \includegraphics[width=0.3\linewidth]{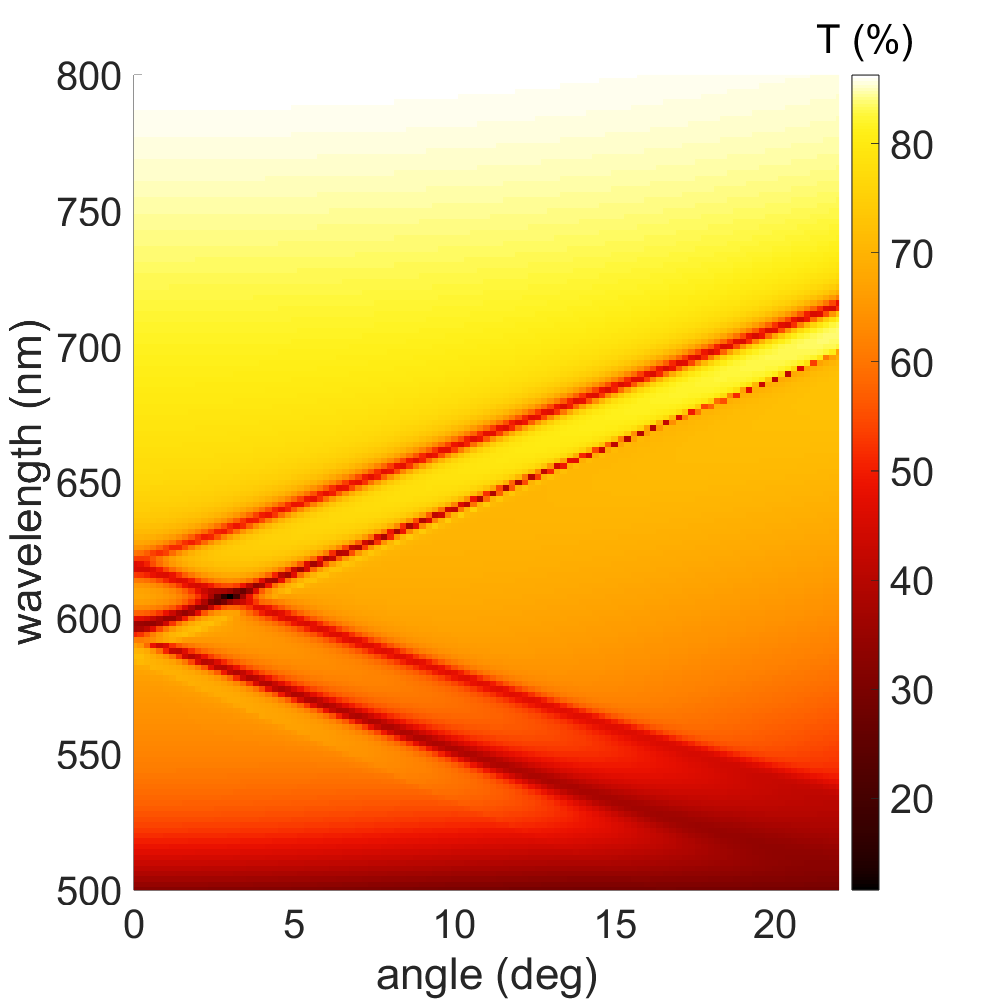}
    \includegraphics[width=0.3\linewidth]{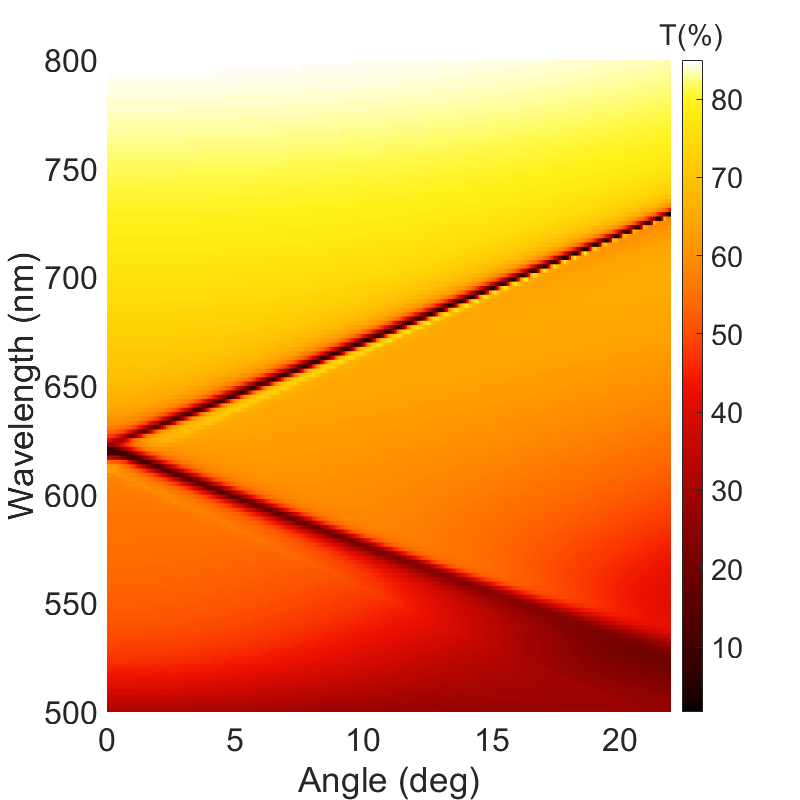}
    \includegraphics[width=0.3\linewidth]{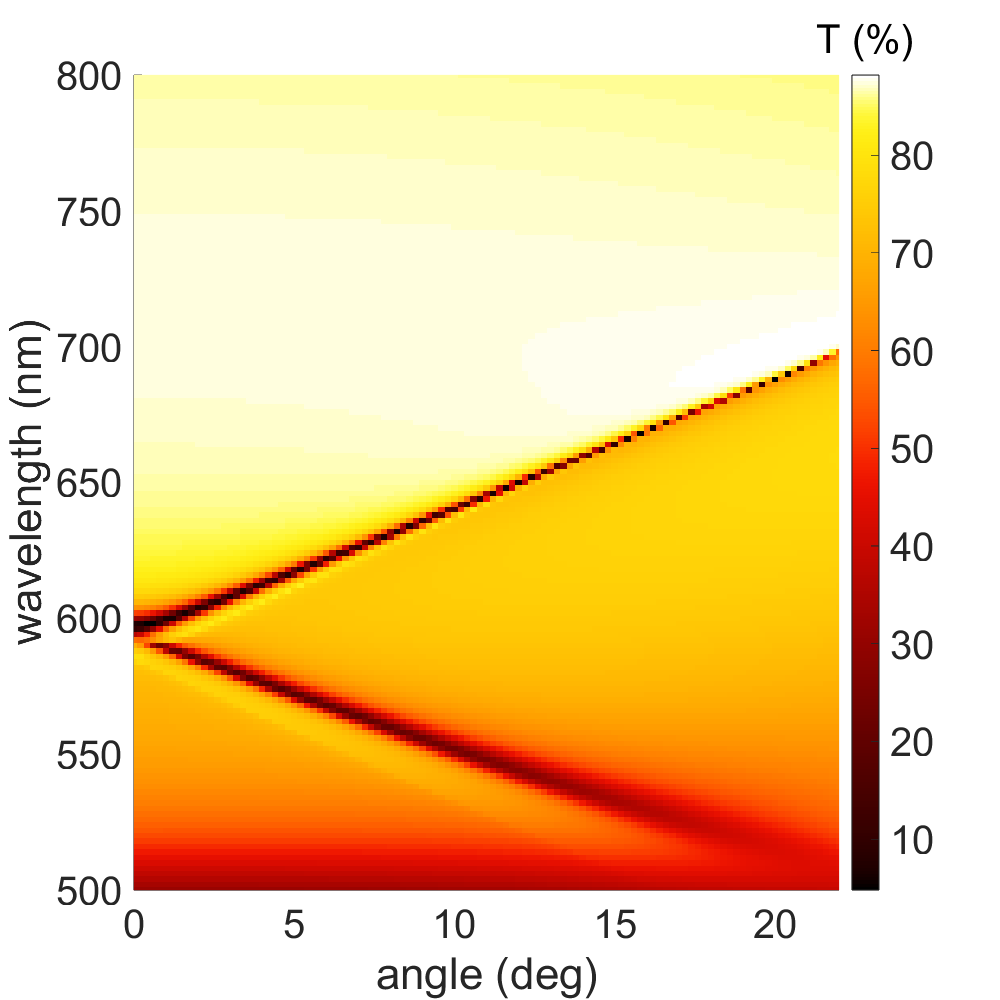}\\
     \textbf{d}~~~~~~~~~~~~~~~~~~~~~~~~~~~~~~~~~~\textbf{e}~~~~~~~~~~~~~~~~~~~~~~~~~~~~~~~~~\textbf{f}\\
     \includegraphics[width=0.3\linewidth]{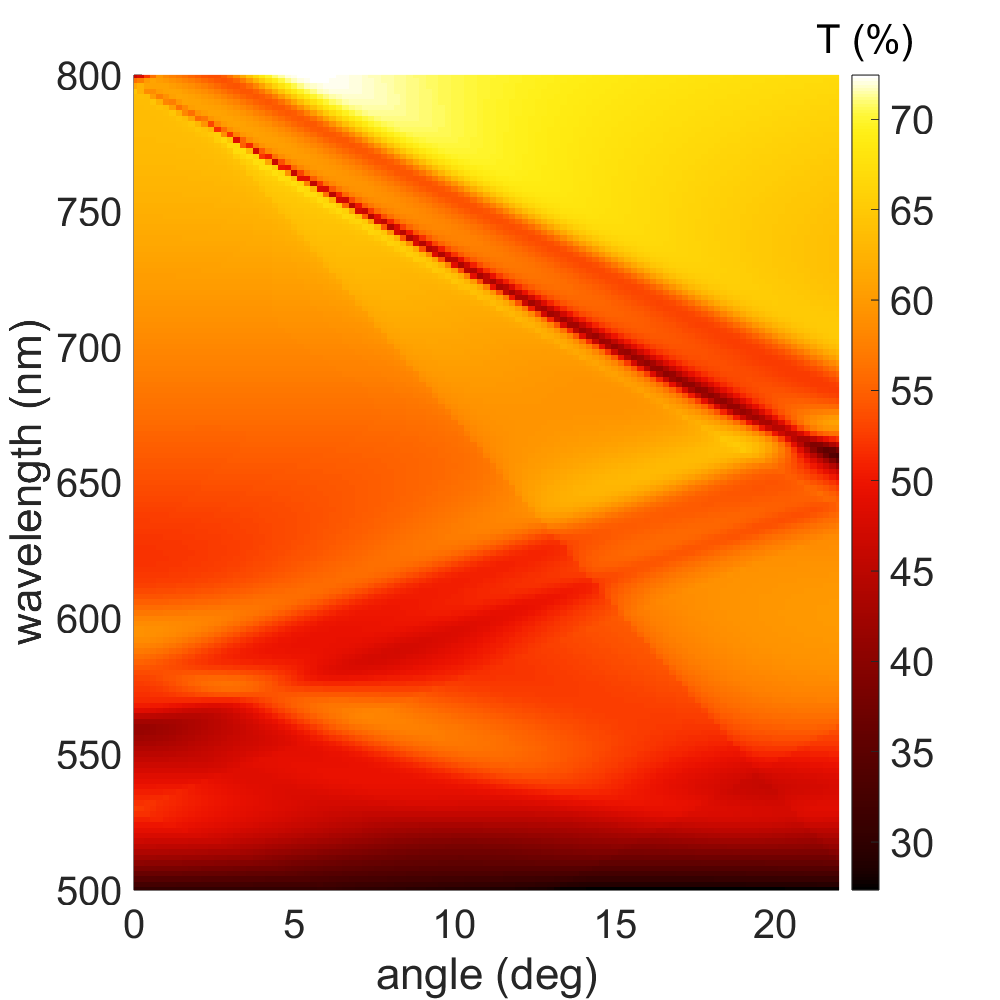}
    \includegraphics[width=0.3\linewidth]{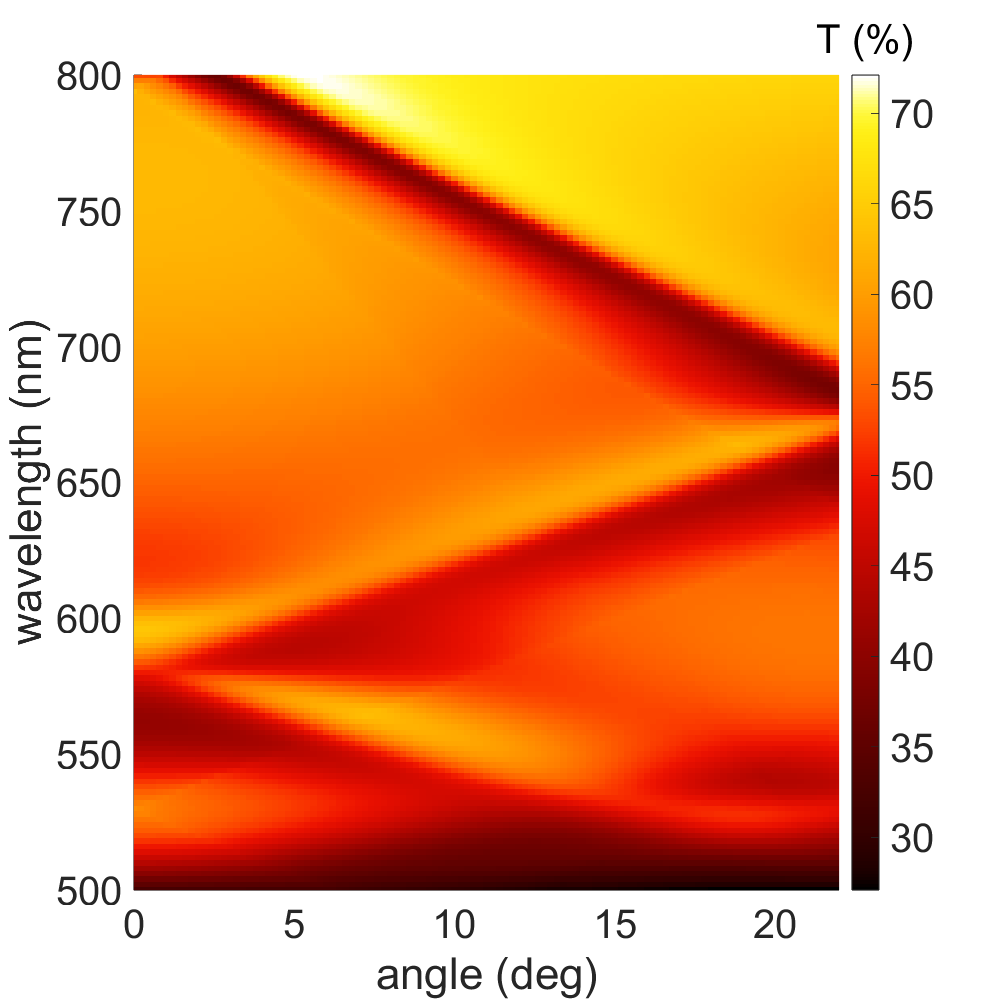}
        \includegraphics[width=0.3\linewidth]{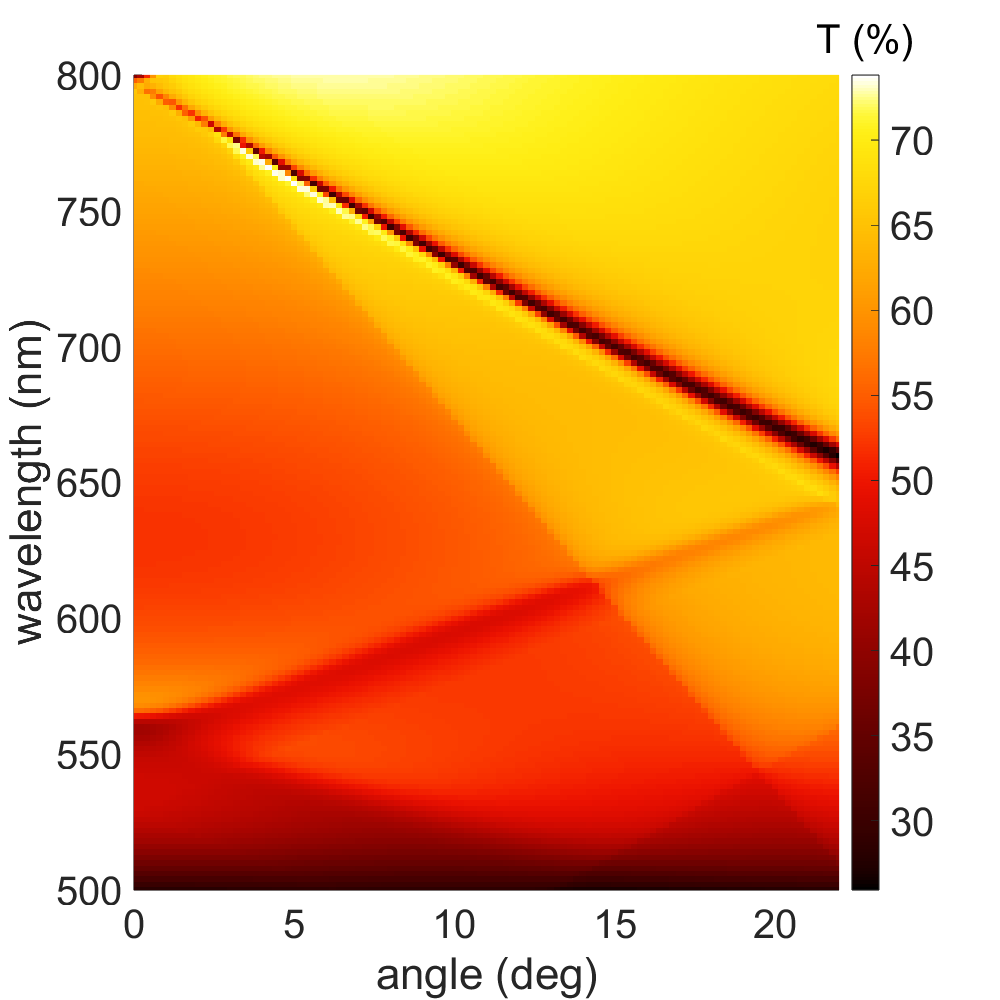}
    
    \caption{The numerically simulated transmittance spectra of the waveguide gratings with 
 $P=300$~nm (top pane, \textbf{a-c}) and $P=800$~nm (bottom pane, top pane, \textbf{d-f}) calculated for the circularly polarized (left column, top pane, \textbf{a,d}), s-polarized (central column, \textbf{b,e}) and p-polarized (right column, \textbf{c,f}) incident light.}
    \label{Fig: Model spectra}
\end{figure}

The rigorous coupled wave analysis method (RCWA) \cite{moharam1995formulation, li2003fourier} was used to calculate the optical response of structured multilayer media, each layer of which is either uniform or periodic in one or two directions. The method is based on the solution of Maxwell's equations in a truncated Fourier space. Special factorization rules are used to achieve good convergence. Maxwell's equations take the form of an eigenvalue problem to find the Bloch waves of each layer. Then the eigenfunctions are combined taking into account the boundary conditions. In this way, it is possible to calculate the distributions of electric and magnetic fields, as well as the coefficients of reflection, transmission, and so on. The available software implementation makes it possible to take into account optical anisotropy as well as the magneto-optical properties of materials.

We performed the numerical simulations of the waveguide gratings using the following values of dielectric permittivity and gyration of
BIG, GGG and $\mathrm{TiO}_2$: $\epsilon_\mathrm{TiO_2} = 5.1819$, $\epsilon_\mathrm{BIG} = 4.8566+0.0183i$, $g_\mathrm{BIG} = 0.0117$, $\epsilon_\mathrm{GGG} = 3.8$ in the vicinity of the light wavelength $\lambda\approx700$~nm. The calculated optical spectra (see Fig.~\ref{Fig: Model spectra}) are in a very good agreement with the experimentally obtained ones (see Fig.~\ref{Fig: Methods Grating spectra}). This proves the consistence of the model used for simulations and justifies the correctness of the near-field distributions of the electomagnetic fields and profiles of the optomagnetic effects inside the waveguide grating.




\section*{Declarations}


\begin{itemize}
\item Funding Basis Foundation Russian Federation (22-2-2-44-1)
\end{itemize}

\bibliography{_Bibliography}

\end{document}